\documentclass[twocolumn]{aastex631}

\usepackage{pifont}

\usepackage{mathtools}

\usepackage{graphicx}
\usepackage{amsmath}
\usepackage{natbib}
\usepackage{amssymb}
\usepackage{hyperref}
\usepackage{xcolor}

\newcommand{\pv}{\left(\Gamma\beta\right)_{\rm sh}}
\newcommand{\pvnot}{\left(\Gamma\beta\right)_{\rm sh,0}}
\newcommand{\pvell}{\left(\Gamma\beta\right)_{\rm sh,\ell}}

\begin{document}
\title{Numerical Modeling of Relativistic Effects in Synchrotron-Emitting Shocks}

\correspondingauthor{Ross Ferguson}
\email{fergu734@umn.edu}

\author[0009-0004-9306-291X]{Ross Ferguson}
\affiliation{School of Physics and Astronomy, University of Minnesota, Minneapolis, MN 55455, USA}

\author[0000-0001-8405-2649]{Ben Margalit}
\affiliation{School of Physics and Astronomy, University of Minnesota, Minneapolis, MN 55455, USA}

\begin{abstract}
Synchrotron emission is seen in a vast array of astrophysical transients, such as gamma-ray bursts (GRBs), radio supernovae, neutron star (NS) mergers, tidal disruption events (TDEs), and fast blue optical transients (FBOTs). Despite the ubiquity of synchrotron-emitting sources, modeling of the emergent flux from these events often relies on simplified analytic approximations for the radiative transfer. These approximations are inaccurate for high-velocity shocks, where special-relativistic effects are important. Properly incorporating these effects considerably complicates calculations, and generally requires a numerical treatment. In this work we present a novel numerical model which solves the full radiative-transfer problem in synchrotron-emitting shocks, accounting for all relativistic effects. This `full-volume' model is capable of calculating synchrotron emission from a shock of arbitrary velocity, and is designed to be flexible and applicable to a wide range of astrophysical sources. Using this new code, we evaluate the accuracy of more commonly-used approximate models. We find that the full-volume treatment is generally necessary once the shock proper-velocity exceeds $\pv \gtrsim 0.1$, and that approximate models can be inaccurate by $\gtrsim$ an order-of-magnitude in trans-relativistic shocks. This implies that there may be a bias in the inferred physical properties of some FBOTs, jetted TDEs, and other relativistic explosions, where approximate analytic models are typically employed. The code associated with our model is made publicly available, and can be used to study the growing population of relativistic synchrotron-emitting transients.
\end{abstract}


\section{Introduction}\label{intro}

A wide-ranging set of transient astrophysical sources give rise to strong collisionless shock waves propagating through circumstellar media (CSM). As such shocks sweep up material, they can accelerate electrons to relativistic velocities. Due to the presence of magnetic fields behind the shock front, themselves generated by plasma instabilities,  the relativistic electrons emit synchrotron radiation. This phenomenon is well-observed in various classes of transients, such as radio supernovae, gamma-ray bursts (GRB), and tidal disruption events \citep[e.g.,][]{Weiler86,Chevalier98,Sari98,Bloom11,Burrows11,Berger14}.

A common assumption is that some fraction of CSM electrons are accelerated by the shock into a non-thermal power-law energy distribution by a first-order Fermi process. The precise mechanism that produces such a distribution in a post-shock fluid is thought to be diffusive shock acceleration \citep{Bell,BO78,BE87}, an idea that has theoretical support from particle-in-cell (PIC) simulations \citep[e.g.,][]{Spitkovsky08,SironiSpitkovsky09,SironiSpitkovsky11,Park11}. Power-law distributions have been successful in describing observations of synchrotron radiation in many events, including non-relativistic radio supernovae \citep{Chevalier82,Weiler86, Chevalier98,ChevalierFransson17} and ultra-relativistic GRB afterglows \citep{Sari98,GS02,vanEerten12,Fong15,Ryan20}. 

A non-thermal power-law electron population is not the only 
possible source of synchrotron emission. It is expected that most downstream electrons are thermalized, with PIC simulations indicating that only a relatively small fraction of the shock energy goes into a non-thermal power-law electron distribution \citep[e.g.,][]{Park11,Crumley19,Jikei25}. Synchrotron emission from thermal electrons has been considered in some specific cases, such as accreting supermassive black holes and GRB afterglows \citep[e.g.,][]{Ozel,GianniosSpitkovsky09,ResslerLaskar17,Warren22}, but is often neglected in broader contexts.

Prior theoretical work indicates that thermal electrons are important for modeling emission from mildly relativistic shocks with proper velocity $\pv \sim 1$ \citep{Margalit&Quataert21, MQ24}, where the proper velocity of the shock is related to the shock velocity and Lorentz factor via $\beta_{\rm sh} =\pv \big/{\sqrt{1+\pv^2}}$ and $\Gamma_{\rm sh} = \sqrt{1 + \pv^2}$, respectively. The trans-relativistic regime $\pv \sim 1$ appears to be important for a variety of observed sources. Modeling of Fast Blue Optical Transients (FBOTs) infers a range of shock velocities $\beta_{\rm sh} \sim 0.1 -0.5$ that approach this regime \citep[e.g.,][]{Margutti19,Ho19,Coppejans20,Ho20,Yao22,Ho2021,Nayana25}. Jetted Tidal Disruption Events (TDEs) also appear to possess shocks with relativistic velocities \citep[e.g.,][]{Bloom11,Burrows11,Giannios&Metzger11,Andreoni22,Rhodes23,Matsumoto23}. \cite{Ho2021} and \cite{Rhodes25} find evidence for the presence of synchrotron emission from thermal electrons in the FBOT AT2020xnd and the jetted TDE AT2022cmc, respectively. 
In contrast, \cite{Nayana25} have recently modeled synchrotron emission from the FBOT AT2024wpp and find no evidence for emission by thermal electrons in this event.
Resolving these tensions and shedding further light on the micro-physics at play will require systematic and accurate modeling of emission from trans-relativistic explosions.

The modeling of synchrotron emission produced by trans-relativistic shocks has a variety of applications in addition to FBOTs and jetted TDEs. Synchrotron-emitting shocks have been shown to be important in the context of neutron star (NS) mergers \citep[e.g.,][]{Nakar11,Piran13,Hotokezaka15,Margalit15,Margalit20,Kathirgamaraju2019,Nedora21,Nedora23}. An accurate exploration of the synchrotron emission can help constrain the physics of such mergers, especially considering the possibility of magnetar-boosted emission from NS mergers \citep{Metzger14,Fong16,Horesh16,Liu+20,Schroeder20,Ricci+21,Sarin22} and given the wealth of data on the merger GW170817 \citep[e.g.,][]{Mooley18,Alexander18,Hajela22}.
Relativistic shock waves interacting with CSM material can also occur in broad-line Type Ic supernovae \citep{Kulkarni98,Corsi16,Sun25} and low-luminosity GRBs \citep[e.g.,][]{Nakar12,BarniolDuran15,Hamidani25}. Deeper analysis of the resulting synchrotron emission may help test models for the progenitors of these events.

With the above cases in mind, there is a clear need for accurate models of synchrotron emission and self-absorption in cases where the shock has a proper velocity $\pv \sim 0.1-10$.
Unfortunately, as we explain below, existing theoretical models used for modeling generic events are typically inaccurate in this trans-relativistic regime because they do not properly account for all of the relativistic effects that modify observed emission (see however \citealt{Sadeh23,Sadeh24_B} for recent calculations including such effects in the specific context of kilonova afterglows). The goal of our present work is to remedy this situation by presenting a new formalism that fully treats these relativistic effects without any approximations to the radiative transfer. This formalism can be applied to arbitrary shock velocities and hydrodynamic solutions. We treat the hydrodynamics as input to be specified by users, and expect that the particular choice will vary depending on astrophysical context. For the sake of concreteness, in this study we select a simple toy-model prescription for the hydrodynamics that allows us to isolate the effects of relativistic time-of-flight and Doppler boosting on the radiative transfer calculation.

The emergent synchrotron flux produced by a shock, including the effects of synchrotron self-absorption, can be calculated by solving the radiative transfer equation along all sight-lines (rays) from the source to the observer. 
In general, this involves numerical integration over a series of ordinary differential equations (ODEs).
Typically, this arduous process is bypassed in favor of simplified analytic approximations  \citep[e.g.,][]{Chevalier98,Sari98,MQ24}. Such analytic approaches can be categorized as  `effective line-of-sight (LOS) approximations',  whereby emission and absorption are estimated along a characteristic ray and the total flux is extrapolated from this single estimate. While this approach is often reasonably accurate for non-relativistic shocks, it is often very inaccurate when relativistic effects become important.  

In GRB afterglow models, the literature evolved from such analytic approaches \citep[e.g.,][]{Paczynski93,Meszaros97,Sari98,Rhoads99} to numerical models that take into account emission from the full emitting volume without resorting to approximations \citep{Granot+99a,Granot+99b,GS02,vanEerten12,Ryan20,Wang24}. The goal of this work is similar: we present code that solves the full radiative transfer problem without approximations, taking into account all relativistic effects. In contrast to the analogous work on GRB afterglows \citep{Granot+99a,Granot+99b}, our model is not limited to cases where the velocities are ultra-relativistic or where the hydrodynamics follow the Blandford-McKee solution. Instead, our model is flexible and applicable to arbitrary velocities, including the trans-relativistic $\pv \sim 1$ regime relevant to transients such as FBOTs, jetted TDEs, NS mergers, broad-lined Ic supernovae, and low-luminosity GRBs.

In addition to GRBs, prior work has considered synchrotron emission from mildly relativistic ejecta launched as a consequence of compact mergers \citep{Sadeh23,Sadeh24_A,Sadeh24_B}. These studies include the effects of ejecta on the system's evolution, and find analytic approximations which match the calculations of detailed numerical calculations. Our model, in contrast, allows the user to freely specify the hydrodynamic solution and can be used to model synchrotron emission in generic contexts. In the present formulation, our model utilizes a simple hydrodynamic solution which ignores the presence of ejecta and any post-shock fluid evolution.

As we show in this work, previous approximate approaches can differ significantly  from the exact results of our numerical model. In particular, the combination of moderately relativistic velocities, non-homogeneous CSM density profiles, and decelerating shocks lead to a significant underestimation of the peak flux and the optically thin behavior of the emission. 

The paper is structured as follows: In \S\ref{sec:approximations}, we present an overview of the different approaches to calculating the emergent synchrotron flux, in particular comparing the full-volume method to the approximate methods utilized in later sections. In \S\ref{model}, we develop the details of the full-volume model, including the hydrodynamics of the shock, the post-shock physics, and the radiative transfer problem. In \S\ref{model_results}, we explore the model's predictions for different choices of shock velocity, CSM density, and deceleration and compare the results to approximate approaches. In each case, we construct spectral energy distributions (SEDs) and contour plots showing where emission originates behind the shock. In \S\ref{sec:spatial_distribution}, we consider how the emission is spatially distributed behind the shock and explore in what cases the region diametrically opposed to the observer (the `distal region') is important to the emergent flux. \S\ref{sec:curve_fit} fits the full-volume model and the effective LOS approximation to data of the FBOT CSS161010 \citep{Coppejans20} in order to examine the different parameter estimates in each case. In \S\ref{sec:discussion} we summarize our model and discuss caveats and future directions. 

\section{Overview of Radiative Transfer: full-volume and Approximate Methods}\label{sec:approximations}

The total flux from a synchrotron-emitting source, including the effects of emission and self-absorption, are calculated via radiative transfer. However, the difficult task of solving the full radiative transfer equation is often circumvented by the use of various approximations. In this work, we consider three methods for calculating the synchrotron flux, which can broadly be thought of as one-dimensional (the effective LOS approximation), two-dimensional (the thin shell approximation), and three-dimensional (the full-volume model). Only the full-volume method calculates the emergent flux without recourse to any approximations.

The simplest reasonable approach to solving the radiative transfer problem is a one-zone model, here referred to as an `effective LOS' approximation. 
Instead of formally solving the radiative transfer ODE, it is assumed that the emission coefficient $j_\nu$ and absorption coefficient $\alpha_\nu$ have some characteristic values in an emitting region of width $\Delta R$. In this approach, the emergent intensity can be approximated as $I_\nu \sim (j_\nu/\alpha_\nu) [ 1 - \exp(-\alpha_\nu \Delta R )]$ \citep{RBL}. The luminosity can then be estimated by assuming that $I_\nu$ is representative of all sight-lines, so that $L_\nu\sim 4\pi(\pi R^2) I_\nu$, with the characteristic radius $R$ related to the size of the emitting region such that the emitting surface area visible to the observer is $\sim\pi R^2$. Typically, the width of the emitting region is also related to this characteristic radius as $\Delta R \propto R$.

The characteristic values of $j_\nu$, $\alpha_\nu$, and $R$ are evaluated at some reference position which is assumed to be representative of the shock properties as a whole. Using the hydrodynamics described in \S\ref{model}, we find that this assumption can be reasonable for certain types of non-relativistic explosions, yet quickly breaks down once relativistic effects become important. Note that the degree to which the effective LOS model is accurate in non-relativistic explosions can depend strongly on the hydrodynamics. In this work, we isolate the impact of relativistic effects on the effective LOS approximation by assuming a simple hydrodynamic solution such that the post-shock fluid properties do not change after interacting with the shock wave.

As an illustrative example, consider Doppler corrections. Doppler boosting significantly enhances or inhibits emission depending on the Doppler factor $D$, which varies substantially throughout the emitting region, since it depends on the fluid proper velocity $(\gamma\beta)_f$ and on the angle between the fluid velocity and the observer LOS. The observed emission from different regions of the shock can therefore vary by orders of magnitude when $\pv \gg 1$. That is---there is no single reference position that is representative of the entire shock.

Nevertheless, it is still possible to write down one-zone models that include a one-zone treatment of relativistic effects \citep[e.g.,][]{Sari98,Giannios&Metzger11,MQ24}. {\it If} the emergent emission from the whole shock is actually dominated by emission from one particular region, {\it and if} the one-zone model is chosen such that the considered ``zone'' corresponds to this same region, this approach may still yield reasonable approximations of the true emission. However, this is generally not the case, and as we show later in this work, simple one-zone models can be wildly inaccurate even for trans-relativistic shocks.

The advantage of this class of models is that they are fully analytic. 
In fact, nearly all existing analytic models of synchrotron emission from transients fall under this category of one-zone effective LOS models \citep[e.g.,][]{Chevalier82,Chevalier98,Sari98,Giannios&Metzger11,Nakar11,Margalit&Quataert21,MQ24}. 
These types of models are the de-facto standard in the field and are used extensively in the literature (with the notable exception of GRB afterglows, where more detailed models are widely available and commonly used; e.g., \citealt{GS02,vanEerten12,Ryan20}).

\begin{figure*}
 \centering
  \includegraphics[trim=4cm 6cm 0.5cm 6cm,clip=true,width=0.9\textwidth]{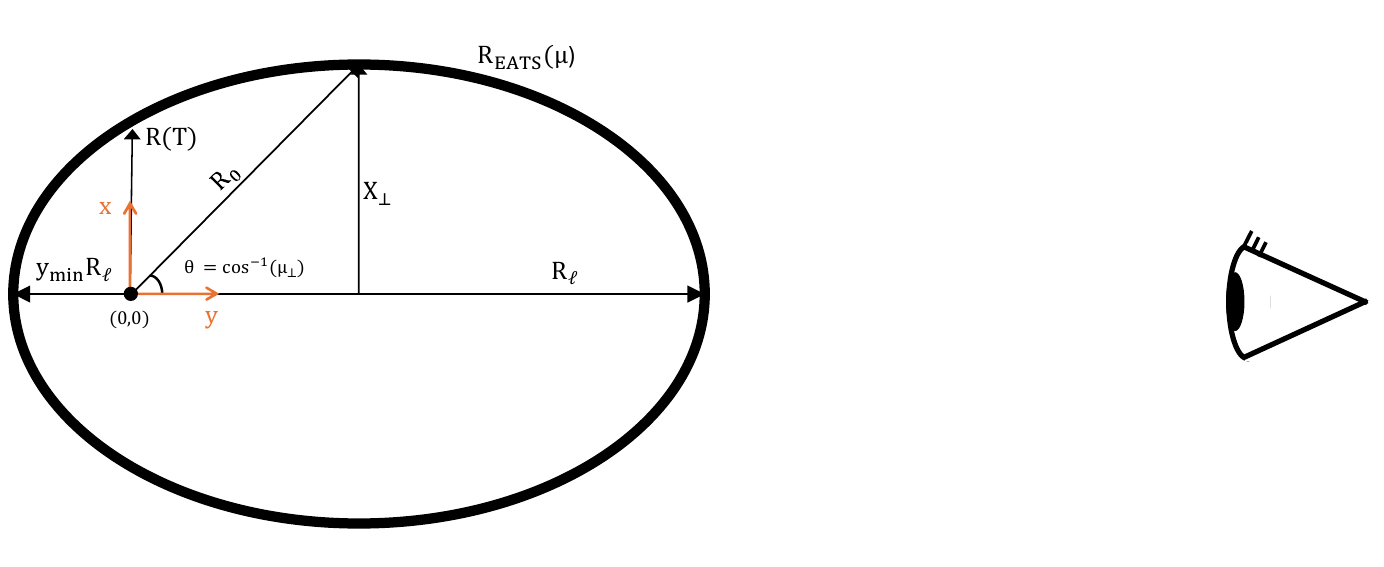}
\caption{Schematic diagram showing the cross section of a relativistic shock, with the observer located to the right. The equal arrival-time surface (EATS) $R_{\rm EATS}$ encloses the volume from which photons can be emitted behind the shock and arrive contemporaneously at time $T$. 
At any given epoch, an observer sees emission from (potentially a subset of) particles contained within this region.
Note that $R_{\rm EATS}(\mu,T)$ is a function of angle from the origin $\mu = \cos \theta$, and that the
corresponding oblateness of this surface is purely a consequence of photon time-of-flight effects. That is, the shock itself is assumed to be perfectly spherical, but relativistic effects cause it to {\it appear} oblate.
The degree of oblateness is correlated with the shock Lorentz factor, $X_\perp / R_\ell \sim  1/\Gamma_{{\rm sh},0}$,
where $R_\ell$ and $X_\perp$ denote the greatest possible physical distances parallel and perpendicular to the observer, respectively (Equations~\ref{R_l}, \ref{X_perp}). The radius corresponding to $X_\perp$ is labeled $R_0$ (Equation~\ref{eq:R0}), whereas $R(T) \equiv R(t=T)$ is the shock radius for which $y=0$. The smallest y-coordinate on the EATS lies along the LOS in the direction opposite of the observer, and is labeled $y_{\rm min}$. 
}
  \label{fig:EATS}
  \end{figure*}

A particularly important one-zone approach is a LOS approximation, in which the reference position is chosen to be at the shock front, exactly along the LOS (corresponding to an angle $\theta=0$; see Figure~\ref{fig:EATS}). In this work, we instead use an \textit{effective} LOS approximation where the reference position is chosen to be somewhere other than the LOS. In \S~\ref{model_results}, we choose this position to be the maximum perpendicular position of the shock. Note that \cite{MQ24} use a different reference position based on a simple prescription of the shock dynamics. In \S~\ref{sec:curve_fit}, we examine two effective LOS models, one which employs the maximum perpendicular position as in \S~\ref{model_results} and the other which uses the \citealt{MQ24} radius. 

As a potential improvement on these approximate one-zone models
we also consider a
two-dimensional `thin-shell' approach, whereby emission and absorption are assumed to be coming from an infinitesimally thin shell just behind the shock front. This two-dimensional surface is then assumed to be representative of the fluid properties, a generalization of the one-zone model. 
Unlike the effective LOS approximation, different rays do have different absorption and emission properties, and we must numerically integrate over the contribution from each. 
However, the radiative transfer equation for each ray can be solved analytically under the thin-shell approximation, circumventing the need to numerically solve a large number of ODEs. The approach can therefore be categorized as semi-analytic, requiring only a single numerical integration for each flux calculation. This method therefore has the benefit of being more accurate than effective LOS models while remaining computationally inexpensive. 
A detailed description of the thin-shell formalism is provided in Appendix~\ref{appendix:thin_shell}; see also \cite{Wang24} for a different implementation of a thin shell approximation .

The thin shell approximation is particularly valid for optically thick frequencies, when observed emission originates very close to the shock surface, and for non-relativistic shocks. This model has an advantage over the effective LOS approximation in that it can account for a range in Doppler factors as well as relativistic time-of-flight effects that alter the shape of the shock front as seen by an observer (discussed in \S~\ref{model:kinematics}; see also \citealt{Granot+99a}). 
However, as we later show, the predictions of this model still fall short of the full (three-dimensional) un-approximated calculation in most other circumstances.
Nevertheless, the thin shell model is useful as a comparison point to the full-volume calculation, and it is often more accurate than more commonly used effective LOS models.

The three-dimensional `full-volume' approach is described in detail in \S\ref{model} and closely follows models that were first developed by \citealt{Granot+99a,Granot+99b} for GRB afterglows (albeit our formalism introduces important extensions to these works). This approach takes into account relativistic effects and local changes to fluid properties without making simplifying assumptions regarding emission and absorption. The radiative transfer equation is solved numerically for a given ray, and contributions to the intensity from different rays are integrated over as in the thin shell approximation. The full-volume model is more cumbersome and computationally intensive than the one- and two-dimensional approximations, but it is the only model that properly calculates emission without any approximations while accounting for all relativistic effects. Several sections of this paper are devoted to exploring in which circumstances the machinery of the full-volume model is necessary and in which a simpler approximation is sufficient. As we will see, accurate predictions in generic contexts require the full-volume approach.

\section{General Formalism}\label{model}

\subsection{Time-of-Flight Effects}\label{model:kinematics}
We describe the motion of a shock using the shock radius $R(t)$, defined in terms of the model-dependent proper velocity $\pv(t)$ discussed in \S\ref{sec:downstream}. In this section we explore the consequences of an arbitrary choice of $\pv$ on the hydrodynamics, following the general formalism of \cite{Granot+99a,Granot+99b}. 

We assume spherical symmetry and that the hydrodynamic variables can be written as
functions only of time and the self-similar radial coordinate $\xi = r/R(t)$. Defining $\mu = \cos{\theta}$, we use a spherical coordinate system $\{r,\mu,\phi\}$ centered at the origin of the explosion and align the $\mu = 1$ axis along the LOS.

The photons seen by an observer at time $T$ were emitted at different retarded times $t = T + \mu r/c$; due to the $\mu$-dependence an observer sees a shock that appears distorted from its spherical shape. The Equal Arrival Time Surface (EATS) from which emitted photons reach the observer at identical times is described by a contour $R_{\rm EATS}(\mu,T)$. Given the shock expansion history $R(t)$ we can obtain $R_{\rm EATS}$ implicitly via
\begin{equation}\label{R_max1}
R_{\rm EATS}(\mu,T) \equiv R \left( \left. t \right\vert_{{\rm EATS}} \right)  = R\left( T + \frac{\mu R_{\rm EATS}}{c}\right).
\end{equation}

In order to solve the implicit Equation~(\ref{R_max1}) the expansion history of the shock must be specified.
For the purposes of the present study, we consider a simple parameterization in which the shock proper-velocity scales as a power-law in $R(t)$ (see Equation~\ref{prop_vel} below). A complete analysis of this transformation is given in Appendix~\ref{app:shock_radius} assuming that the power-law deceleration scaling holds for all times $t>0$. In that appendix, we show that $R_{\rm EATS}$ can be written implicitly in terms of hypergeometric functions. In  Figure~\ref{fig:EATS}, we illustrate an example EATS for a mildly relativistic shock. 

Using $R_{\rm EATS}(\mu,T)$, we can find the maximum shock radius $R_\ell$ parallel to the LOS and the maximum perpendicular distance $X_{\rm \perp}$ from the LOS to the EATS,
\begin{equation}\label{R_l}
R_\ell \equiv R_{\rm EATS}(\mu = 1, T),
\end{equation}
\begin{equation}\label{X_perp}
X_{ \perp} \equiv \max_{\mathbf{\mu}} \hspace{2pt} \left[ \sqrt{1-\mu^2} R_{\rm EATS}(\mu, T) \right].
\end{equation}
We define the angle which satisfies Equation~(\ref{X_perp}) to be $\mu_{\rm \perp}$ and the EATS radius at that point as 
\begin{equation} \label{eq:R0}
    R_0 \equiv R_{\rm EATS}(\mu_\perp,T) .
\end{equation}

For the hydrodynamics considered in this paper, we find the simple relation $\mu_\perp = \beta_{\rm sh,0}$ (Equation~\ref{eq:mu_perp}), where $\beta_{\rm sh,0}$ is the shock velocity at time $t_0$ when the shock radius satisfies $R(t_0) = R_0$.
Note that
\begin{equation}\label{X_perp_estimate}
    X_\perp = \sqrt{1-\mu_\perp^2} R_0 = R_0/\Gamma_{\rm sh,0},
\end{equation}
so that, in the non-relativistic and ultra-relativistic limits where $R_\ell/R_0$ is independent of velocity (see Equation~\ref{RlR0}), the oblateness of the EATS can be characterized as $X_\perp / R_\ell \propto  1/\Gamma_{{\rm sh},0}$. A non-relativistic shock with $\pv \ll 1$ has $X_{\rm \perp}\simeq R_\ell$, so the EATS is approximately spherical. On the other hand, for an ultra-relativistic shock with $\pv\gg 1$, $X_{\rm \perp}\ll R_\ell$ and the shock becomes greatly stretched along the LOS.

We introduce non-dimensional Cartesian coordinates 
\begin{equation}\label{y}
y \equiv \frac{\mu r}{R_\ell},
\end{equation}
\begin{equation}\label{x}
x \equiv \frac{r \sqrt{1-\mu^2}}{X_{\rm \perp}}.
\end{equation}
To preserve consistency with the \cite{Granot+99a} formalism, the coordinates are chosen such that $y$ is aligned along the LOS and $x$ is perpendicular to it. The definition above implies that $x\in[0,1]$, so that we look at only the upper half-ellipse in Figure~\ref{fig:EATS}. The $\mu$-dependence of the shell gives the EATS only azimuthal symmetry, despite the spherical expansion of the shock.  Due to the assumption of spherical symmetry, emission and absorption for all azimuthal angles $\phi$ is identical.

Figure~\ref{fig:EATS} pictures two key radii not yet mentioned. The first is the the minimum EATS distance $y_{\rm min}R_\ell$, which is located diametrically opposite to the observer along the LOS. The second is the distance $R(T)$ from the origin to the EATS for $y=0$ has no time-of-flight delay ($t = T$); this radius separates emission from the region facing the observer ($y>0$, $R_{\rm EATS} > R(T)$) and emission from the region facing away from the observer ($y<0$,$R_{\rm EATS} < R(T)$). We define these emitting volumes to be the `proximal' and `distal' regions, respectively. The relative importance of each---assuming emission from the distal region is not inhibited by possible intervening ejecta---is considered in \S\ref{sec:spatial_distribution}.

\subsection{Radiative Transfer}\label{subsec:rad_transfer}

We now consider synchrotron emission and self-absorption from both power-law and thermal electrons behind the shock. The power-law electrons are injected into a distribution $(\partial n/\partial\gamma) \propto \gamma^{-p}$ with spectral index $p>2$. In contrast, the thermal electrons are described by a relativistic, thermal Maxwell-J$\ddot{\textnormal{u}}$ttner distribution. For both electron populations, we  adopt the emission and absorption coefficients from \cite{Margalit&Quataert21} (Equations 10, 12, 14, and 16 in that work). These equations must be evaluated in the local fluid rest frame (primed quantities) at Doppler-shifted frequencies $\nu' = \nu/D(\gamma_f,\mu)$. The Doppler factor can be written in terms of the fluid proper velocity $(\gamma\beta)_f$ as
\begin{equation}\label{Doppler}
D(\gamma_f,\mu) 
= \frac{1}{\gamma_f \left( 1- \mu \beta_f \right)}
= \left[ \sqrt{1+(\gamma\beta)_f^2} - \mu (\gamma\beta) _f\right]^{-1} .
\end{equation}
In the present work, we neglect the effects of synchrotron cooling and inverse-Compton scattering. To obtain the flux and luminosity of the shock, we proceed as follows. Fixing a value of $x$, we use the radiative transfer equation   ${dI_\nu}/{ds} =j_\nu - \alpha_\nu I_\nu$ to numerically solve for the integrated intensity at the front of the shock. In the local fluid rest frame, this becomes
\begin{equation}\label{dI_dy}
 \frac{1}{R_\ell}\frac{dI_{\nu}}{dy} = D^2(\gamma_f,\mu)j_{\nu'}(y,t) - \frac{1}{D(\gamma_f,\mu)} \alpha_{\nu'}(y,t)I_{\nu}(y).
\end{equation}

Solving Equation~(\ref{dI_dy}) for different values of $x$ yields the specific intensity $I_\nu(x)$ at the front of the shock. Note that this implicitly assumes that scattering is negligible, which is generally true in our context, since the Thompson optical depth is $\tau_{\rm T} \sim \sigma_{\rm T} n_e R_\ell \ll 1$. To obtain the specific flux, we must integrate over $I_\nu(x)$. Using the perpendicular area element $dA_{\perp} = 2\pi X_\perp^2 x dx$, the flux is \citep{Granot+99b}
\begin{equation}\label{F_nu}
F_\nu \simeq \frac{1+z}{d_L^2} \int dA_{\perp} I_\nu = 2\pi (1+z)\bigg(\frac{X_\perp}{d_L}\bigg)^2 \int_0^1 x I_\nu (x) dx,
\end{equation}
where $d_L$ is the luminosity distance and $z$ is the redshift.
Finally, the isotropic equivalent luminosity can be calculated as $L_\nu = 4\pi d_L^2 F_\nu$. 

The general solution to the radiative transfer is independent of the specific prescription for the hydrodynamic evolution, local fluid variables, and emission and absorption coefficients. In the next section, we describe the choices made for these functions in the present work. In Appendix~\ref{sec:Appendix_GRB}, we compare our results to the conventions used for ultra-relativistic blast waves launched in GRB afterglows, for which a detailed full-volume model was developed in \cite{Granot+99a}, \cite{Granot+99b}, and \cite{GS02}. For this purpose, we change the hydrodynamics and radiation variables to those of the Blandford-McKee solution and standard GRB conventions, and find excellent agreement between the full-volume framework and models used for GRB afterglows. This serves as a code-validation test for our work.

\subsection{Model-Dependent Assumptions}\label{sec:downstream}

Sections~\ref{model:kinematics},\ref{subsec:rad_transfer} lay out the general formalism of our `full-volume' model. This formalism can be applied to arbitrary settings, 
and Equations~(\ref{dI_dy}, \ref{F_nu}) can be solved to give the emergent flux $F_\nu$
so long as the 
shock expansion history $R(t)$,
the post-shock fluid velocity profile $(\gamma\beta)_f(r,t)$,
and the emissivity $j_\nu(r,t)$ and absorption coefficients $\alpha_\nu(r,t)$ are specified.
In the present section, we describe the specific choices adopted for these quantities in our present work. It is important to note that these choices are not an inherent part of our full-volume formalism, and are only made for the sake of concreteness and convenience. Our general methodology and its associated code can be applied with alternative prescriptions for any of these variables, as may be relevant in different astrophysical contexts.

For the expansion history of the shock $R(t)$, we assume a `power-law deceleration' model, the details of which are derived in Appendix~\ref{app:shock_radius}.
Within this model, we parameterize deceleration by assuming that the shock proper velocity $(\Gamma\beta)_{\rm sh}$ scales as a power-law with the shock radius,
\begin{equation}\label{prop_vel}
\pv = \pvnot \bigg(\frac{R(t)}{R_0}\bigg)^{-\alpha} .
\end{equation}
Here $\pvnot$ is a nominal value for the shock proper velocity, 
corresponding to the shock velocity at the benchmark radius $R_0$ (chosen here to be the maximum perpendicular radius; see Figure~\ref{fig:EATS})
and the index $\alpha \geq 0$ controls shock deceleration. 
$\alpha = 0$ corresponds to a non-decelerating shock that coasts at a constant velocity, while $\alpha > 0 $ describes a shock that slows down as it sweeps through the ambient CSM.
The shock radius $R(t)$ at the time of photon emission is fully determined by the values of $\pvnot$ and $\alpha$. An implicit expression for $R(t)$ is given in Equation~(\ref{RLOS2}).

For the CSM density profile,
we assume that the external CSM number density $n_{\rm ext}$ takes the power-law form
\begin{equation}\label{n_ext}
n_{\rm ext}(r) 
= n_0 
\left( \frac{r}{R_0} \right)^{-k},
\end{equation}
where $n_0$ is a nominal value for the density
(evaluated at the benchmark radius $r=R_0$; see above)
and the index $0 \leq k < 3$ controls the steepness of the density profile. 

In the non-relativistic and ultra-relativistic limits, a simple closure relation connects $k$ and $\alpha$ for an energy-conserving point explosion. A spherically symmetric adiabatic blast wave expanding into a power-law density medium has total energy 
\begin{equation}\label{explosion_energy}
E \sim \Gamma_{\rm sh}(\Gamma_{\rm sh}-1) M_{\rm sw}c^2 \propto \Gamma_{\rm sh}(\Gamma_{\rm sh}-1) R^{3-k},
\end{equation}
where $M_{\rm sw} \propto \int^R_0 n_{\rm ext}(r) r^2 dr$ is the mass swept-up by the explosion. In both the non-relativistic and ultra-relativistic limits,
an energy conserving solution is satisfied when
$\pv \propto R^{-(k-3)/2}$. For the power-law deceleration model parameterized following Equation~(\ref{prop_vel}), 
this implies the
closure relation $\alpha = (3-k)/2$. This relation describes the Sedov-Taylor and Blandford-McKee solutions, which are valid in the non-relativistic and ultra-relativistic regimes, respectively \citep{Sedov46,Taylor50,BlandfordMcKee76}. 

In the trans-relativistic regime, this closure relation is no longer strictly valid, and a simple power-law deceleration does not ensure energy conservation. Even so, relativistic corrections to the Sedov-Taylor solution are of only second order in $\beta_{\rm sh}$, and may therefore be relatively small \citep{Coughlin19}. 
Note also that the closure relation above only holds for shocks that are in the energy conserving self-similar phase of their evolution. At early times, the shock dynamics are additionally governed by the ejecta of material launched in the initial explosion. This generally leads to smaller values of $\alpha$ than implied by the energy-conserving expression, down to $\alpha \approx 0$ in the very early ejecta-dominated (coasting) phase of the explosion \citep[see, e.g.,][]{Truelove&McKee99}.
For these reasons we leave $\alpha$ as a free parameter and do not restrict our application to values obeying the non/ultra-relativistic adiabatic closure relation discussed above.

For the post-shock hydrodynamic variables, we adopt a simplistic model in which these variables are taken to be spatially constant behind the shock. 
With this prescription, the electron number density, fluid energy density, and the fluid Lorentz factor are assumed to have no radial dependence between 
$\xi_{\rm shell}\leq\xi\leq1$, where $\xi \equiv r/R(t)$ is a self-similar radial coordinate. For $\xi<\xi_{\rm shell}$, we assume that there is no emitting or absorbing material. The value of $\xi_{\rm shell}$ is set by the requirement that all of the CSM material swept-up by the shock is confined within this region (Appendix~\ref{app:xi_shell}).
In this model, the hydrodynamic variables are determined entirely by the shock jump-conditions.

We assume an effective adiabatic index $\hat{\gamma} = (4+\gamma_f^{-1})/3$, so that the fluid proper velocity---as determined by the shock jump conditions \citep{BlandfordMcKee76}---can be written as 
\begin{equation}\label{GB_f}
\left(\gamma \beta\right)^2_{f} = \frac{1}{4} \left[ \pv^2 - 2 + \sqrt{\pv^4 + 5\pv^2 + 4} \right] .
\end{equation}
Similarly, the downstream number density as measured in the fluid rest frame is (Equation~\ref{n_prime2})
\begin{equation}
    n' = 4\gamma_f n_{\rm ext} \left( R[t] \right)
    ,
\end{equation}
where $\gamma_f = \sqrt{ 1 + \left(\gamma \beta\right)^2_{f} }$.
The corresponding
electron number density is $n_e' = \mu_e n'$, where $\mu_e$ the mean molecular weight per electron. 
Finally, the downstream energy density takes the form
\begin{equation}\label{u_prime}
u' = (\gamma_f-1) n' \mu_u m_p c^2,
\end{equation}
where $\mu_u$ is the mean molecular weight.

We adopt emission and absorption coefficients following \cite{Margalit&Quataert21}, which account for synchrotron emission and absorption from an electron distribution function that contains both thermal and non-thermal (power-law) electron populations.

These coefficients depend on the local fluid Lorentz factor $\gamma_f$, number density $n'$, energy density $u'$, dimensionless electron temperature $\Theta = kT_e/m_e c^2$, and magnetic field $B$. 
We assume that the post-shock magnetic field, power-law electrons, and thermal electrons each constitute constant fractions of the total energy density:  $u_B = \epsilon_B u'$, $u_{e} = \epsilon_e u'$, $u_T = \epsilon_T u'$. The magnetic field $B$ in the post-shock fluid rest frame is therefore 
\begin{equation}\label{B}
B = \sqrt{ 8\pi \epsilon_B u' } ,
\end{equation}
and the effective electron temperature $\Theta$ can be similarly calculated by setting $\epsilon_{T} u' = a(\Theta) n_e' \Theta m_e c^2$, with the function $a(\Theta)$  approximately given by $a(\Theta) \simeq (6+15\Theta)/(4+5\Theta)$ \citep{Gammie,Ozel,Margalit&Quataert21}.

The expressions in \cite{Margalit&Quataert21} implicitly assume that the majority of post-shock electrons reside within the thermal population, which is the case in the limit $\epsilon_e/\epsilon_T \ll 1$. This assumption is motivated by the results of PIC simulations, and is consistent with the fiducial parameters adopted in the present work. Nevertheless, we note that this assumption can easily be relaxed by multiplying any occurrence of $n_e$ in \cite{Margalit&Quataert21} by a correction factor $\left[1 + (\epsilon_e/\epsilon_T) a(\Theta) \Theta /( \frac{p-1}{p-2} \gamma_m -1 ) \right]^{-1}$.
Without this correction, electron number and kinetic energy are not strictly conserved. However, for the parameters explored in the current work, this correction never exceeds a few percent, and we neglect it here.

The simple model explored in this paper has four key parameters describing the physical dynamics: the benchmark proper velocity $\pvnot$, the power-law deceleration index $\alpha$, the benchmark CSM density $n_0$, and the power-law density index $k$. These parameters are related to the functional forms chosen for $\pv(t)$ (Equation~\ref{prop_vel}) and $n_{\rm ext}(t)$ (Equation~\ref{n_ext}).

In addition, the model requires the specification of the downstream microphysics. The current implementation includes six microphysical parameters: the equipartition parameters $\epsilon_e$, $\epsilon_B$, and $\epsilon_T$; the power-law index $p$; and the mean molecular weights $\mu_e$ and $\mu_u$.\footnote{In the literature, the mean molecular weights are both typically assumed to be unity; in this work we instead choose solar values. These parameters are left out of any curve fitting.} 
There is a significant amount of degeneracy in the equipartition parameters whose origin lies in the poorly constrained post-shock microphysics. For the purpose of fitting to data, the equipartition parameters are typically considered fixed parameters. Further considerations of parameter fitting are taken up in more detail in \S\ref{sec:curve_fit}.

\section{Comparison to Approximate Models}\label{model_results}
 \begin{figure*}
 \centering
  \includegraphics[width=1.0\textwidth]{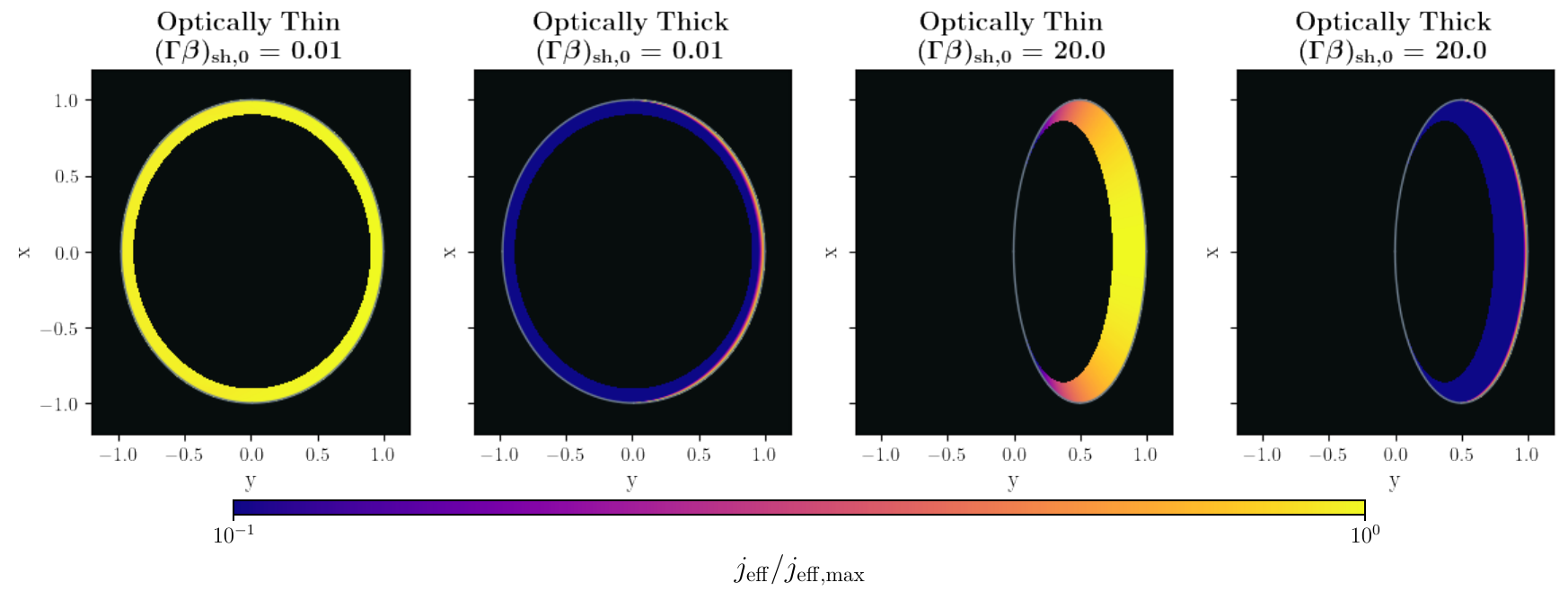}
  \caption{
  Emission region contour plots for constant shock velocity and CSM density. These plots show contours of the `effective emissivity' $j_{\rm eff}$, defined via Equation~(\ref{eq:j_eff}). The effective emissivity is normalized here to the maximum value in each plot, with a log-scaled colorbar extending from 0.1 to 1.0. This effective emissivity is a measure of how much each region contributes to the total emergent flux at frequency $\nu$, accounting for both Doppler corrections and synchrotron self-absorption. The total luminosity is proportional to the volume integral over this quantity ($L_\nu \propto \int j_{\rm eff} \,dV$). The left two panels correspond to a non-relativistic shock with proper-velocity $\pv = 0.01$, while the right two panels are for a highly-relativistic shock with $\pv = 20$. The first and third panels are computed at optically thin frequencies, while the second and fourth are in the optically thick part of the spectrum. These plots show that: (i) Observed emission in the optically-thick regime originates from a very thin region near the rightwards facing surface of the EATS (in the direction of the observer); (ii) In contrast, in the optically-thin regime all particles contained within the EATS volume contribute to observed emission; 
  (ii) In the relativistic case, the shape of the EATS is no longer spherical, and observed emission is dominated by the front (right-hand) side of the shock due to Doppler boosting. }
  
  \label{fig:high_low_contour}
  \end{figure*}

   \begin{figure*}
 \centering
  \includegraphics[width=1.0\textwidth]{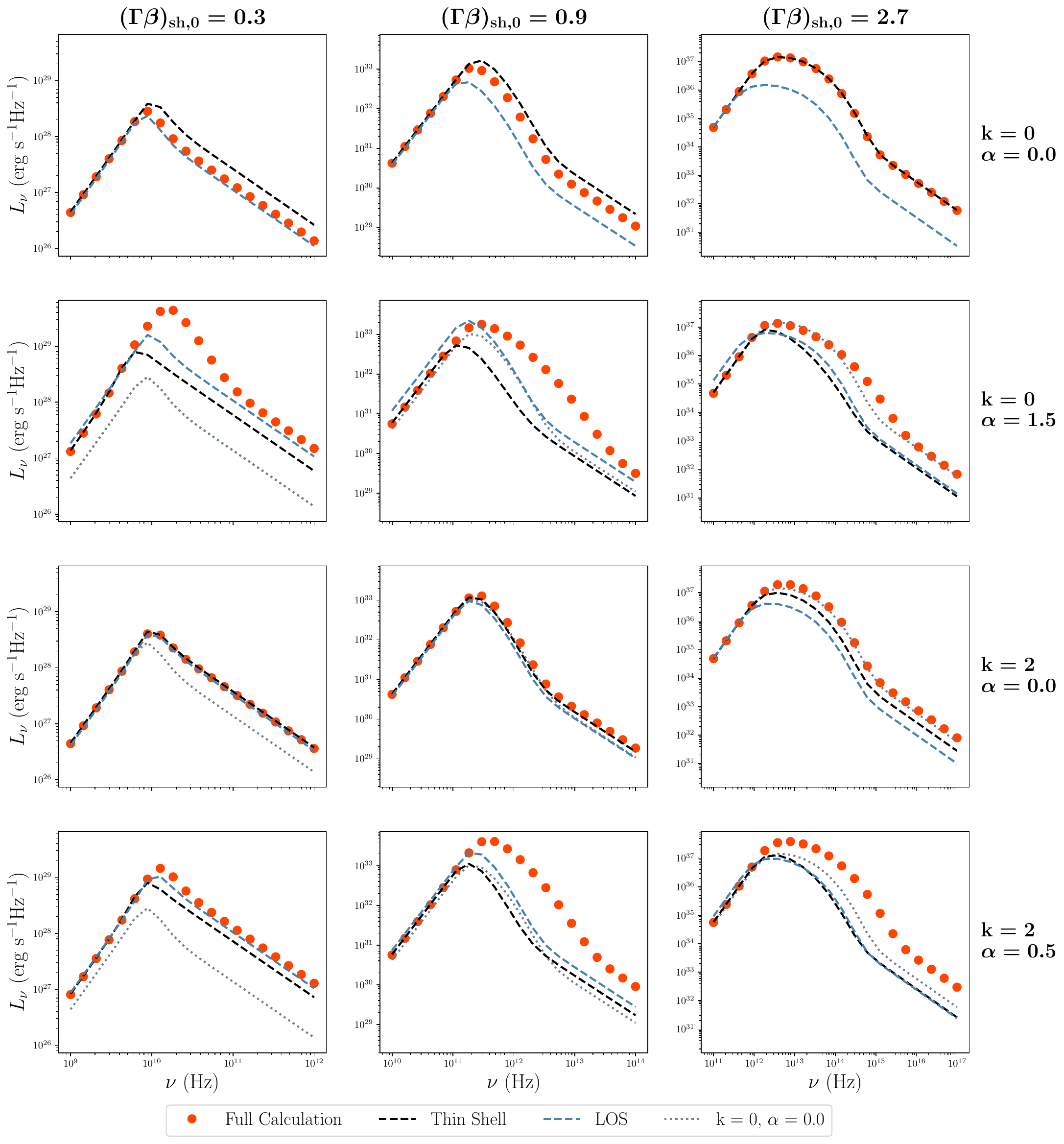}
  \caption{Broadband spectral energy distributions (SEDs) comparing shocks with proper velocities $\pvnot = 0.3$, $0.9$, and $2.7$ (corresponding to the three columns) between models with different deceleration ($\alpha$) and density ($k$) profiles. Models with a given proper velocity are plotted four times (different rows), each time with a different pairing of $k$ and $\alpha$ (labeled on the right-hand side).
  SEDs predicted by the thin-shell (dashed black) and effective-LOS approximations (dashed blue) are plotted alongside the full-volume calculation (orange dots). The $k=0,\alpha=0$ case is shown in each plot for comparison purposes (dotted gray). 
  The discrepancy between the full-volume calculation and approximate models indicates that
  relativistic radiative transfer effects are important even in the trans-relativistic regime, where simple approximations do not reliably describe the full structure of the SED. In certain cases, the predicted fluxes can disagree by over an order of magnitude, and appear morphologically different (e.g., middle panel second row).
  Note that the absolute frequency and luminosity values in these plots are somewhat arbitrary, and depend sensitively on the fiducial parameters, particularly CSM density and observed time.
  }
  \label{fig:SED}
  \end{figure*}

In this section, we explore the physical consequences of the model described in \S\ref{model} and compare its predictions to those of the thin shell and effective LOS approximations. For this purpose, we assume fiducial values $\epsilon_e=0.01$, $\epsilon_T=0.4$, $\epsilon_B=0.1$, $p=3$, $\mu_e=1.18$, $\mu_u=0.62$, $z=0$, $T=50\rm$ days, $d_L=10^{28}\hspace{2pt}\rm cm$, and $n_0 = 10^3 \hspace{2pt}\rm cm^{-3}$. Note that these choices are arbitrary and do not affect our analysis here. The final three model parameters---$\alpha$, $\pvnot$, and $k$---are left free. As we will see, the approximations generally fail to accurately predict the behavior of the shock emission. To see why this is the case, we further examine how much each region behind the shock contributes to the total flux by defining an effective emissivity
\begin{equation}\label{eq:j_eff}
j_{\rm eff}
\equiv
j_{\nu'}(x,y) D^2(x,y) e^{-\tau_{\nu'}(x,y)}, 
\end{equation}
where $D(x,y)$ is the Doppler factor (Equation~\ref{Doppler}) and $\tau_{\nu}(x,y) = R_\ell \int_1^y dy'' \alpha_{\nu'}(x,y'')/D(x,y'')$ is the optical depth from the observer to the point $y$. At each point in the integral, $\alpha_{\nu'}$ and $D$ are calculated at $t = T + \mu r/c$, so that $\tau_\nu$ depends on absorption through the changing downstream medium. 
The effective emissivity $j_{\rm eff}$
represents the contribution 
of point $(x,y)$ to the emergent specific flux observed at time $T$ and frequency $\nu$.
As such, the total luminosity is simply proportional to the volume integral of the effective emissivity $L_\nu \propto \int j_{\rm eff}(x,y) dV$, where $dV = 2\pi R_\ell X_{\perp}^2  x dx dy$.
In the optically-thick case, points far behind the shock would have $\tau_{\nu'} \gg 1$ and contribute very little to the emergent flux (even if locally emission $j_{\nu'}$ at these points is strong). In the optically-thin case $\tau_{\nu'} \ll 1$ and $j_{\rm eff}$ simply tracks the local emissivity, modulated by Doppler corrections.

There are two cases where the thin shell approximation is valid: when the emission is centered at points near the EATS and when time-of-flight effects are unimportant. In the latter case, the spatial extent of the emission region is unimportant, so the assumption that the emission can be squished into a thin shell is reasonable. For the effective LOS approximation to work, we must choose an average radius which is representative of the emission from the shock as a whole. In this section, we choose the characteristic radius to be $R_0$. The average radius weighted by emission may give more accurate results, but is unknown a priori and would require calculating the full-volume model. Furthermore, it turns out that no single radius is representative of the emission in generic cases (see \S~\ref{sec:spatial_distribution}).
  
Figure~\ref{fig:high_low_contour} shows contour plots of $j_{\rm eff}$ (emission region contour plots) for shock velocities $\pvnot = 0.01$ and $\pvnot = 20$ at frequencies in the optically thin and optically thick parts of the spectrum. In the optically thick cases, the approximations are accurate: the emission comes from a very thin shell just behind the shock front and $R_0$ is a reasonable average radius. In the optically thin case, the low velocity emission region is centered at a single radius and time-of-flight effects are unimportant, and the approximations are accurate here as well. On the other hand, the high-velocity optically thin case has a thicker emitting region and non-representative emission at $R_0$, and the approximations become incorrect. 

\label{sec:model_output}
 \begin{figure*}
 \centering
  \includegraphics[width=1.0\textwidth]{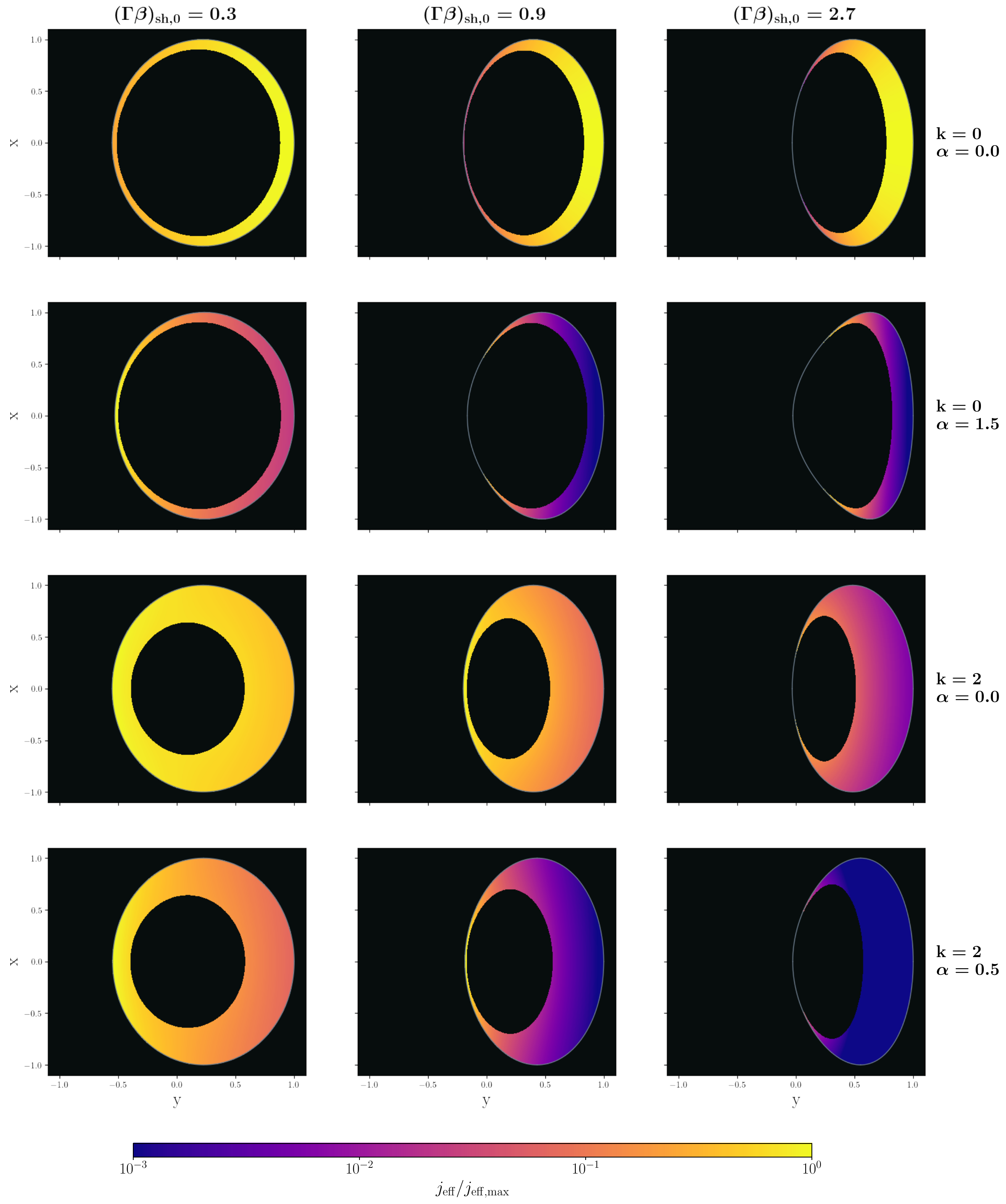}
  \caption{
  Contour plots of the effective emissivity $j_{\rm eff}$ (Equation~\ref{eq:j_eff}) 
  in the optically-thin regime
  for each of the twelve models displayed in Figure~\ref{fig:SED}. The colors show the ratio of effective emissivities in log scale from $10^{-3}$ to $1.0$. Each column represents a characteristic proper velocity $\pvnot$, while each row is calculated for a different pairing of $k$ and $\alpha$ as in Figure~\ref{fig:SED}. The EATS is shown in gray. With a constant velocity shock and a constant CSM density (top row), emission is concentrated towards the front of the shock (right end of the EATS). For decelerating shocks and/or decreasing CSM density profiles ($k,\alpha>0$ respectively), the emissivity becomes stronger in regions towards the back of the shock, and away from the boundary of the EATS ($\xi<1$).
  These facts help explain why the thin-shell and effective LOS approximations can be highly inaccurate, and why the full-volume formalism described in the present work is necessary for reliably calculating emission when $\pvnot \gtrsim 0.1$ (see also Figure~\ref{fig:SED}).
}
  \label{fig:contours}
  \end{figure*}

We now use the framework described in \S\ref{model} to calculate various models at a given observing time $T$ post-shock launch. Twelve example SEDs are shown in Figure~\ref{fig:SED} for three different shock velocities. In Figure~\ref{fig:contours}, we calculate emission region contour plots for the same models. At optically thick frequencies, both approximations tend to agree reasonably closely with the full-volume model; however, for higher frequencies near the peak and in the optically thin regime, the approximations generically become inaccurate. As discussed above, when the shock velocity is pushed to non-relativistic values, the approximations closely agree with the full-volume calculation. In the subsections below, we examine each row in turn.

\subsubsection{Case 1: \texorpdfstring{$k=0$}{k=0}, \texorpdfstring{$\alpha=0$}{alpha=0}}
\label{sec:00plot}

The first rows of Figures~\ref{fig:SED} and \ref{fig:contours} examine non-decelerating shocks ($\alpha=0$) propagating in constant CSM densities ($k=0$). The approximations work well in the optically thick regime; however, at roughly the peak of the SEDs, the optical depth decreases to $\tau\sim1$ and subsequently the approximations become inaccurate. At these shock velocities, the effective LOS approximation underestimates the luminosity, whereas the thin shell approximation overestimates the luminosity. For even higher velocities, as in Figure~\ref{fig:high_low_contour}, both approximations underestimate the luminosity at optically thin frequencies. 
As we can see from Figure~\ref{fig:contours}, the optically thin emission predominantly comes from a shell around the EATS which becomes thicker for higher shock velocity\footnote{As $\pvnot$ increases, the physical width of the emitting region gets smaller ($R\sim 1/\Gamma^2$), but time-of-flight effects make the \textit{apparent} width wider.}. Doppler beaming causes emission to be centered on the front of the shock for higher velocities. 

\subsubsection{Case 2: \texorpdfstring{$k=0$}{k=0}, \texorpdfstring{$\alpha=1.5$}{alpha=1.5}}
\label{sec:01.5plot}
In the second rows of Figures~\ref{fig:SED} and \ref{fig:contours}, we relax the assumption of a constant velocity shock and instead assume the shock decelerates as $\pv \propto R^{-1.5}$. In this case, the effective LOS approximation becomes somewhat inaccurate even in the optically thick part of the spectrum, while the thin shell remains a good approximation in this regime. At optically thin frequencies, both approximations significantly underestimate the luminosity. The optically thin regime in the center plot in this row -- $\pvnot = 0.9$ -- is particularly notable, as not only are the approximations orders of magnitude too low, but the peak frequency and qualitative shape of the SED are different than the approximations.

The emission region contours show a stark departure from the non-decelerating case. The emission now comes primarily from a small region towards the back of the shock. The physical reason for this change is clear: points nearest the origin emit photons at early times when $\pv$ is larger and the emissivities are higher, so emission is centered there. This effect is somewhat moderated for more relativistic shocks by Doppler beaming, and the emission region moves closer to the front of the EATS. Note that for these values of $\pvnot$, $R_0$ is still somewhat representative of the emission, but less so for higher velocity.

\subsubsection{Case 3: \texorpdfstring{$k=2$}{k=2}, \texorpdfstring{$\alpha=0$}{alpha=0}}
\label{sec:20plot}
The third rows of Figures~\ref{fig:SED} and \ref{fig:contours} illustrate the case of a constant velocity shock propagating in a stellar wind CSM ($k=2$). Both approximations again work at optically thick frequencies, but underestimate the luminosity in the optically thin regime. For higher velocities, this underestimation becomes more significant, as the shock quickly reaches regions with low number density. 

For small velocity, the emission regions are only moderately back-heavy and reside in a thicker shell than the constant-density case. As the velocity increases, the effective emissivity becomes clearly centered on the distal region, since these points correspond to earlier times when the shock probed smaller radii with higher CSM densities. In extreme cases, this distal-favoring effect wins over the proximal-favoring Doppler beaming. As discussed in \S~\ref{sec:spatial_distribution}, the total emission is generally still dominated by the proximal region due to the larger volume. For  $\pv\lesssim 0.8$, however, the distal region controls the total emission.

\subsubsection{Case 4: \texorpdfstring{$k=2$}{k=2}, \texorpdfstring{$\alpha=0.5$}{alpha=0.5}}
\label{sec:20.5plot}

The fourth row of Figures ~\ref{fig:SED} and \ref{fig:contours} combines deceleration (slightly lower than in the second row at $\alpha = 0.5$) with a stellar wind CSM density of $k=2$. This pair of $k$ and $\alpha$ corresponds to the approximately energy-conserving solution discussed below Equation~\ref{explosion_energy}. In this model, the SEDs reflect the combination of the previous two cases. Back-heavy emission is created due to both the smaller shock velocity and smaller density at greater radii. Doppler beaming moves the emission more towards the proximal region for higher shock velocities, but this effect is again heavily moderated by deceleration and lower density. As in the previous cases, the back of the shock becomes the strongest source of emission, and thus $R_0$ is not a representative radius. The thin shell approximation is inaccurate in these cases as well, since time-of-flight effects are important and the emission comes not just from the thin shell at the back of the EATS, but from throughout the shocked region (i.e., from $\xi>\xi_{\rm shell}$, but not at the EATS itself).

\section{Spatial Distribution of Emission}
\label{sec:spatial_distribution}

\begin{figure*}
 \begin{center}
  \includegraphics[width=1.0\textwidth,
  keepaspectratio]{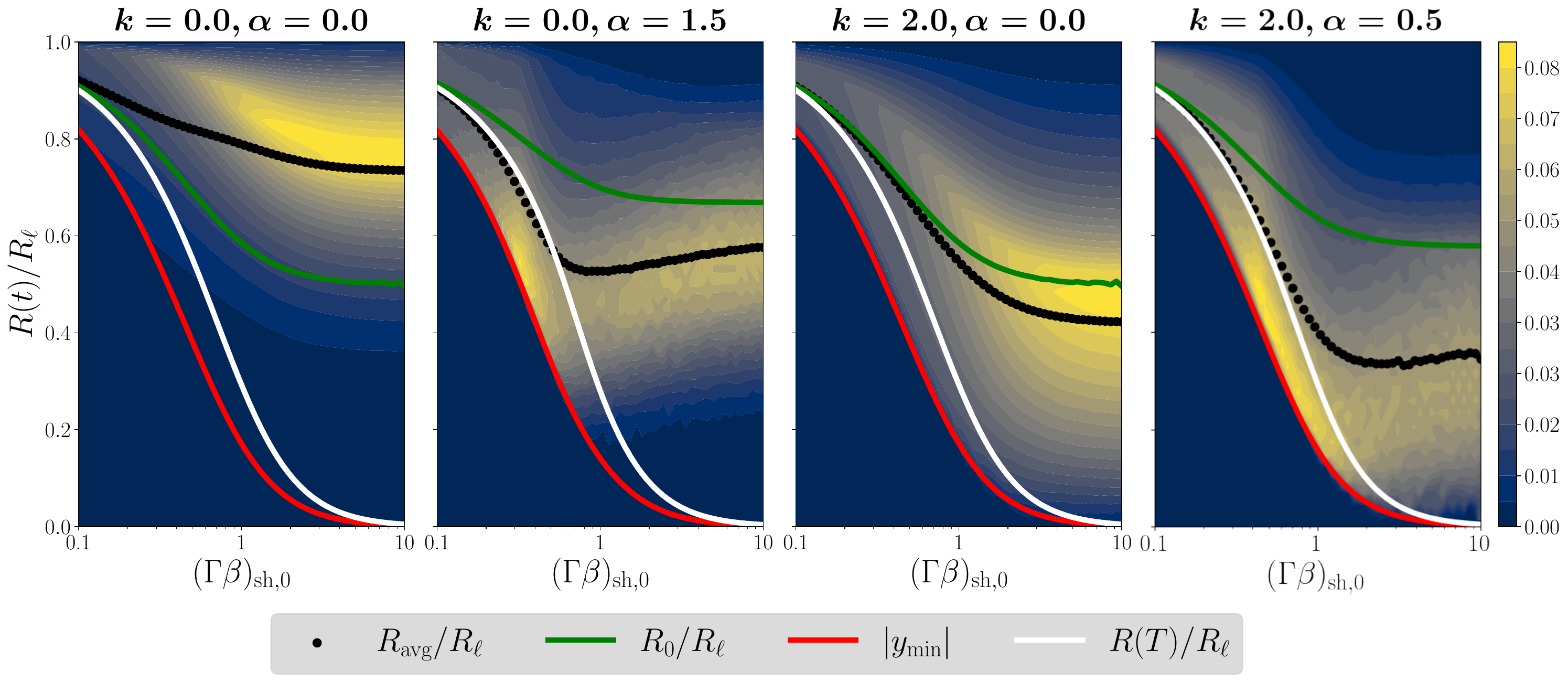}
  \end{center}
  \caption{The spread in shock radius $R$ at various emission times $t$ for four representative models (different values of $k$ and $\alpha$; labeled above each panel) discussed in the text and shown in Figures~\ref{fig:SED}-\ref{fig:contours}.
  At each velocity $\pvnot$, the y-axis displays a histogram of $R(t)/R_{\ell}$ values weighted by the fractional contribution of each bin to the emergent luminosity $L_\nu$ in the optically-thin regime (normalized such that the sum of weights is 1 for each $\pvnot$). 
  These histograms show the range of shock radii probed by contemporaneously-observed emission. For non-relativistic shocks, time-of-flight effects are negligible and there is only a single radius probed at any given time. However, even moderately relativistic shocks with $\pvnot \gtrsim 0.1$ show a significant spread in relevant shock radii. For cases where $k,\alpha \neq 0$, these different radii also correspond to different CSM densities and/or shock velocities that are probed in a single-epoch observation (Equations~\ref{n_ext}, \ref{prop_vel}).
  The class of analytic models typically employed in studying trans-relativistic explosions (effective LOS models) cannot account for this spread.
  For comparison purposes, additional curves are displayed corresponding to $R_0$ (green), the minimum EATS radius $|y_{\rm min}| R_\ell$ (red), the radius $R(T)$ corresponding to $y=0$ (white), and the average radius $R_{\rm avg}$ weighted by the contribution to $L_\nu$ (black points). When the shock decelerates, points in the distal region ($R<R(T)$) control the emission when $\pvnot\lesssim1$.
  }
  \label{fig:r_vals}
  \end{figure*}
  
For relativistic shocks, we must---in principle---consider emission from a large set of radii and times. This fact can most easily be seen by noting that, for large $\pvnot$, the minimum EATS y-coordinate $y_{\rm  min}$ is much less than the maximum EATS y-coordinate $y=1$. More specifically, for $\pvnot \gg 1$, we have $y_{\min}\propto\pvnot^{-2}$ so that $y_{\min} \ll 1$ and the EATS covers a wide range of radii (see Equation~\ref{y_min_scaling} and following discussion). This concept can be recast in terms of the distribution of emission times, since, using $|t-T| = y R_\ell/c$,
\begin{equation}\label{eq:tdist}
\frac{t_{\min} - T} {t_{\max}-T} 
\overset{\pvnot \gg 1}{\propto}
\pvnot^{-2}
\ll
1 .
\end{equation}

On the other hand, due to Doppler boosting, only a much smaller set of radii are truly relevant. For example, photons emitted at time $t_{\min}$ may not contribute appreciably to the emission at optically thin frequencies, as in 
the third panel of Figure~\ref{fig:high_low_contour}. For ultra-relativistic shocks, the range of emitting radii has been explored by other authors (e.g., \citealt{Granot+99a}). In the trans-relativistic regime, the set of relevant emitting radii is not immediately clear.\footnote{
The effect of relativistic distortion on the radii and temperatures of thermally emitting kilonova photospheres has recently been considered in \cite{Sadeh25}.}
Moreover, in several panels of Figure~\ref{fig:contours}, the distal region (related to the counter jet in the context of GRBs), characterized by emission from points for which $y<0$, appears to dominate the overall shock emission, contrary to the expectation that this emission should be Doppler boosted away for relativistic shocks. These cases occur in the presence of deceleration or a stratified density, when the shock front either moved much faster or the external number density was considerably higher for smaller radii (both of which increase the magnetic field and enhance the emission). In this section, we explore the spatial distribution of emission in detail, illustrating in particular when emission from the distal region becomes relevant.

An important factor not seen directly in the emission region contours of Figure~\ref{fig:contours} is the effective volume from the proximal and distal regions. In many cases, despite each point radiating less strongly, the effective emitting volume of the proximal region is so much larger that the distal region, despite appearances, is not dominant. Even so, it turns out that in some trans-relativistic shocks, the distal region can still control the emission.

To quantify the spatial distribution of emission while accounting for the effective emitting volume, we calculate how much different radii contribute to $L_\nu \propto \int \hspace{1pt}j_{\rm eff}\hspace{1pt} dV 
\propto \int \hspace{1pt}j_{\rm eff}\hspace{1pt} x dx dy
\sim \sum x\hspace{1pt}j_{\rm eff}$, with $j_{\rm eff}$ computed on a discretized $(x,y)$ grid. Each point of the grid corresponds to photons emitted at some retarded time $t$ when the shock front was at the radius $R(t)$. Emission from different points then probes conditions $R(t)$ at different times during the shock's evolution, and therefore the correspondingly different upstream densities (Equation~\ref{n_ext}) and shock velocities (Equation~\ref{prop_vel}). 

Figure~\ref{fig:r_vals} depicts the range of shock radii probed in four different cases (corresponding to those presented in Figures~\ref{fig:SED} and ~\ref{fig:contours}), with contours indicating the weighted contribution of different points to $L_\nu$. Note that these calculations are for frequencies deep in the optically thin regime where power-law electrons are dominant. The precise distribution of emission is dependent on frequency and the spectral index. Even so, a qualitative sense can be gained by examining a single optically thin frequency, as we do here. For a given $\pvnot$, the figure can be thought of as a histogram (normalized to one) showing which shock radii (normalized by $R_\ell$) are most relevant for the observed emission, with the effective volume accounted for by the bins. In the non-relativistic limit, time-of-flight effects become negligible and the histogram collapses to a delta-function at $R(t) = R_\ell$ (see, e.g., Equation~\ref{RlR0}). At higher velocities, a wider range of radii is probed at any given observation epoch due to the corresponding range of emission times.

In the first panel, larger radii corresponding to the proximal region dominate for all $\pvnot$, with the distal region Doppler-boosted away at higher velocities. This is evidenced by the fact that $R_{\rm avg} > R(T)$ for all $\pvnot$, where $R_{\rm avg}$ is the emission-weighted average shock radius and $R(T)$ the shock radius that separates the proximal and distal regions (corresponding to $\mu=0$; see Figure~\ref{fig:EATS}). This case is analogous to the third panel of Figure~\ref{fig:high_low_contour}. For the decelerating case in the second panel, $R_{\rm avg} \lesssim R(T)$ when $\pvnot\lesssim 0.5$ and the distal region dominates the overall emission. On the other hand, emission from the distal region is much less important at higher velocities, where $R_{\rm avg} \gg R(T)$. At high $\pvnot$, $R_{\rm avg}$ is noticeably smaller than in the first panel, which indicates that regions with radii much smaller than $R_l$ and $R_0$ are more important when calculating the emission. This trend is also seen quite clearly in the third row of Figure~\ref{fig:contours}. In the third panel, the distal region never dominates the emission, but it remains important in the trans-relativistic regime. The final panel shows perhaps the most realistic model, with the distal region dominating the emission when $\pvnot\lesssim1.1$. Although points at the back emit far more strongly than points towards the front of the EATS even up to $\pvnot=10$, the emitting region is far smaller, and $R_{\rm avg} > R(T)$. In general models, the distal region is of prime importance to the overall emission for trans-relativistic velocities.

\begin{figure}
 \centering
  \includegraphics[width=0.5\textwidth]{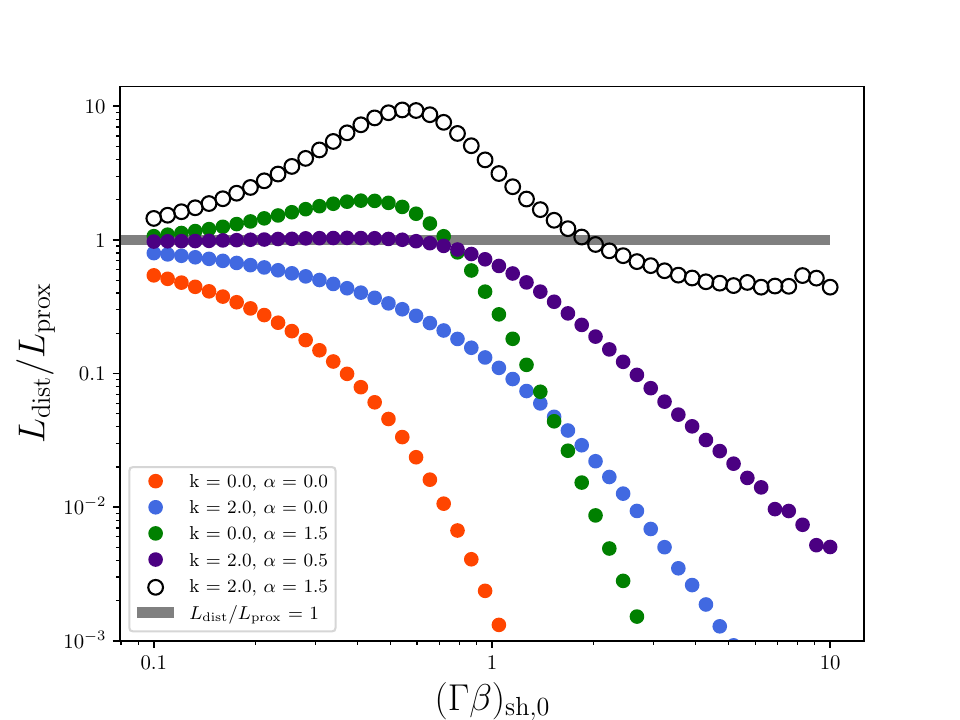}
  \caption{The relative contribution to the emergent luminosity $L_\nu$ from the distal region ($L_{\rm dist}$; region further away from the observer; see Figure~\ref{fig:EATS}) and the proximal region ($L_{\rm prox}$; region where $y>0$) as a function of velocity $\pvnot$, for different pairings of $k$ and $\alpha$. 
  The horizontal gray line delineates between cases where the distal region dominates the total emission (above the gray line) and where it is subdominant (below it). 
  At ultra-relativistic velocities, the combination of Doppler inhibition and the shrinking volume within the $y<0$ portion of the EATS typically cause emission from the distal region to be negligible ($L_{\rm dist}/L_{\rm prox} \ll 1$).
  However, in the trans-relativistic regime $\pvnot \sim 1$, there are cases where 
  $L_{\rm dist}/L_{\rm prox} \gtrsim 1$.
  This implies that emission from the distal region cannot be ignored in trans-relativistic shocks (in comparison to e.g., the \citealt{Granot+99a,Granot+99b} GRB formalism, which neglects this region).
  }
  \label{fig:emission_ratio}
  \end{figure}

We next calculate explicitly whether the distal region or the proximal region emits more of the total luminosity. To do so, we fix an optically thin frequency and define the quantities $L_{\rm dist}$ and $L_{\rm prox}$, corresponding to the luminosity of the distal region $L_\nu(y<0)$ and proximal region $L_\nu(y>0)$, respectively. The ratio $L_{\rm dist}/L_{\rm prox}$ as a function of $\pv$ for various pairings of $k$ and $\alpha$ is shown in Fig~\ref{fig:emission_ratio}. $L_{\rm dist}/L_{\rm prox} < 1$ implies that emission from the proximal region emits more radiation, whereas $L_{\rm dist}/L_{\rm prox}>1$ implies that the distal region dominates. For $\alpha\not=0$, the emission from the distal region either dominates or significantly impacts the total luminosity in the trans-relativistic regime, which can also be observed in Figure~\ref{fig:r_vals}. For extreme values of $k$ and $\alpha$, the distal region is important even for large $\pvnot$; however, staying along the ``energy-conserving" line $\alpha = (3-k)/2$ ensures that for sufficiently high $\pvnot$, the distal emission is Doppler-boosted away. In the extreme, non-energy-conserving case $k=2,\alpha=1.5$ (not examined in Figures~\ref{fig:SED} or ~\ref{fig:contours}), the distal region is quite important even at $\pvnot = 10$, with $L_{\rm dist}/L_{\rm prox} \simeq 1.3$. 

Figures~\ref{fig:r_vals} and \ref{fig:emission_ratio} demonstrate that a wide range of radii must be taken into account when modeling the total emission of realistic shocks in the trans-relativistic regime. The consequences of such a large range of emitting radii are profound: Not only does the distal region become critically important for modeling emission, but the approximations considered in \S~\ref{sec:model_output} become highly inaccurate. The class of one-zone effective LOS models inherently assume that there are single characteristic values for the shock radius, velocity, and external density. Figure~\ref{fig:r_vals} instead shows that there is a wide range of relevant radii with strongly varying local conditions. Though not explicitly shown in Figure~\ref{fig:r_vals}, this corresponds to a large range of CSM densities (if $k>0$; see Equation~\ref{n_ext}) and shock velocities (if $\alpha>0$; see Equation~\ref{prop_vel}) that are probed by emission at a fixed epoch $T$.
Similarly, the thin shell approach cannot provide accurate estimates if much of the emission comes from behind the shock front ($\xi < 1$) and relativistic effects are important. For realistic, moderately relativistic shocks, it is therefore critical to use the full-volume model described above instead of simple approximations.

\section{Implications for Observed Events}\label{sec:curve_fit}
\begin{figure*}
 \centering
  \includegraphics[width=1.0\textwidth]{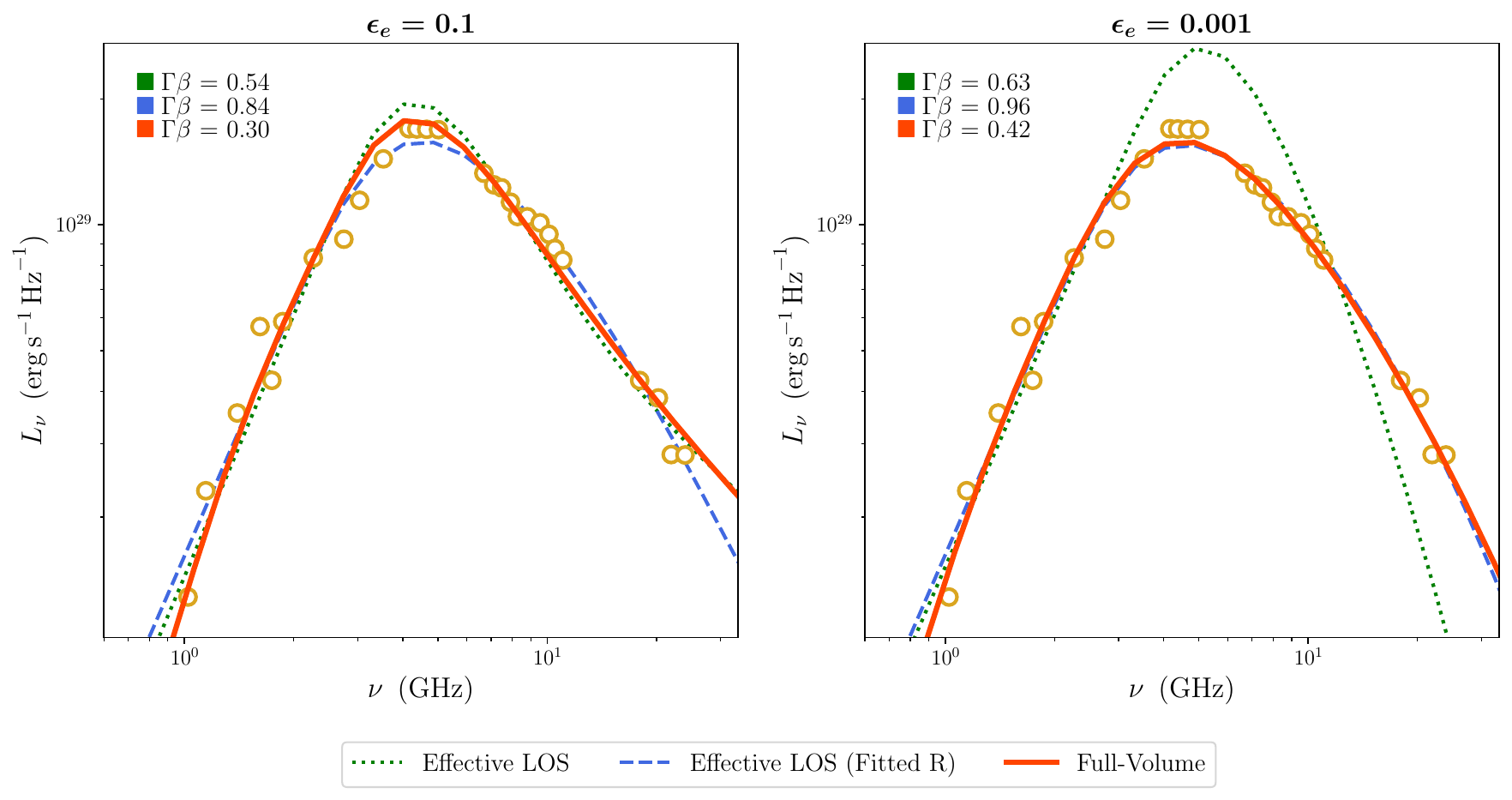}
  \caption{A comparison of the full-volume model and two effective LOS models to data of the FBOT CSS161010 \citep{Coppejans20} at a single epoch ($T = 99\hspace{2pt}\rm days$). In the first panel, we assume that a relatively high energy fraction resides in power-law electrons ($\epsilon_e=0.1$). In the second panel, we instead assume that little energy is deposited into power-law electrons, and instead thermal-electrons dominate the observed SED ($\epsilon_e = 0.001$). The two full-volume models treat the CSM density power-law index $k$ differently: it is assumed to be 0 in the first panel, but is fitted to be $1.96$ in the second. Both full-volume models fit the data well, but the second panel with no power-law electrons captures the high-frequency behavior better. The green curve shows an effective LOS model which assumes an emitting radius $R = \Gamma_{\rm sh}^2 \beta_{\rm sh} cT$, whereas the blue effective LOS model fits for the reference radius. 
  Even though the fitted $R$ LOS model and the full-volume model fit the data well in both cases, their associated inferred parameters differ appreciably (particularly the shock velocity; see top-left of each panel). 
  This discrepancy between the exact full-volume model and approximate methods highlights the importance of using accurate relativistic models in trans-relativistic sources.
  }
  \label{fig:CSS161010_comparison}
  \end{figure*}
To give a sense of how much using different models impacts the inferred parameters in realistic modeling, we compare the predictions of the LOS approximation and the full-volume model when fit to radio data of the FBOT CSS161010 at $T = 99$~days \citep{Coppejans20}. We ignore the measurement uncertainty, as our purpose here is simply to compare different models and not to predict the precise physical parameters of CSS161010. The effective LOS approximation is fit using {\tt scipy.curvefit}, whereas the full-volume model is fit by minimizing the sum of squared error $e = \rm (log_{10}L - log_{10}L_{\rm data})^2$ using the Powell method. 

Two different models are shown in Figure~\ref{fig:CSS161010_comparison}. In each panel, we include the result of the full-volume model (red curve) and two different effective LOS approximations (green and blue curves). The green curve fixes the shock radius to be $R = \Gamma_{\rm sh}^2\beta_{\rm sh} cT$, following the prescription of \cite{MQ24}. We choose this radius so as to compare the full-volume model to a fully analytic effective LOS approximation, without any full-volume calculation being necessary. 

In contrast, the blue curve leaves the radius $R$ as a free parameter that is determined by the fitting routine.

These `effective LOS' and `effective LOS (fitted $R$)' approximations assume the volume-filling fraction in Equation~(\ref{eq:f}), evaluated at $y=1$. The results below depend only weakly on the choice of volume-filling factor.

Both full-volume curves fit for $\pvnot$, $n_0$, $\alpha$, $T$, and either $p$ (first panel) or $\epsilon_T$ (second panel). For reasons given below, the second panel additionally fits for $k$, which is assumed to be $0$ in the first panel. The four LOS curves fit for the same parameters in each panel with the exceptions of $k$ and $\alpha$, which are not defined for the LOS approximation. We include $T$ as a fitted parameter in both full-volume models as a proxy for fitting the shock radius $R_0$ directly. For a particular $\pvnot$ and $\alpha$, and assuming that $\pv\propto R^{-\alpha}$ for a constant value of $\alpha$ throughout the entire evolution of the shock, $R_0$ and $T$ are in one-to-one correspondence given by Equation~(\ref{R0}). In realistic settings, this assumption is not strictly valid during the ejecta-dominated phase. As mentioned in \S\ref{model}, we typically do not fit for the equipartition parameters, as they tend to be highly degenerate---fitting for them with a small amount of data may not yield a unique solution. In the first panel, we fix $\epsilon_e = 0.1$, $\epsilon_B = 0.1$, and $\epsilon_T = 0.4$. In the second panel, we instead fit for $\epsilon_T$ and fix $\epsilon_e = 0.001$ and $\epsilon_B=0.1$. Fitting for $\epsilon_T$ directly is necessary in this case, as this parameter directly controls the thermal electron spectrum which is dominant when $\epsilon_e\ll1$ for the frequencies shown. In this panel, $p$ has been set to $p=2.5$, but this has no effect on the modeling.

The full-volume model fit in the first panel predicts $\pvnot  = 0.305$, $n_0 = 67.19\hspace{2pt}\rm cm^{-3}$, and $p = 2.8$, $\alpha = 1.5$, and $T = 89.1 \hspace{2pt}\rm days$. For these parameters, the sum of squared errors for the full-volume model is $e_1 = 0.08$. The best fit value of $T = 89.1 \hspace{2pt}\rm days$ implies that $R_0$ is smaller than predicted by Equation~(\ref{R0}) at 99 days. This result is not surprising: Equation~(\ref{R0}) assumes that the shock has always been expanding with the fixed power-law profile given in Equation~(\ref{prop_vel}), but in reality we expect $\alpha$ to change as a function of time. Indeed, at early times the explosion is likely ejecta-dominated, with $\alpha\simeq0$. The green fixed $R$ model finds the parameter set $\pvnot = 0.54$, $n_0 = 17.3\hspace{2pt}\rm cm^{-3}$, and $p = 2.44$, while the free $R$ model finds instead finds $\pvnot = 0.84$, $n_0 = 5.35\hspace{2pt}\rm cm^{-3}$, $p = 2.22$, and  $R = 1.07 \times10^{17}\hspace{2pt} \rm cm$. The radius predicted by the fitted $R$ effective LOS model differs slightly from that in the orange full-volume ($R = 1.57\times10^{17}\hspace{2pt} \rm cm$) and green fixed radius ($R = 1.43\times10^{17}\hspace{2pt} \rm cm$) models. The effective LOS curves match the full-volume model decently well, but the predicted physical parameters vary significantly -- particularly the fitted values for $\pvnot$.

The second panel instead assumes that the observed SED is fully dominated by thermal electrons. The fitting for the full-volume model predicts $\pvnot  = 0.418$, $n_0 = 6.25\hspace{2pt}\rm cm^{-3}$, and $\epsilon_T = 0.383$, $k = 1.96$, $\alpha = 1.48$, and $T = 69.72 \hspace{2pt}\rm days$. The introduction of $k$ yields a stronger fit than the first panel---the sum of squared errors is $e_2 = 0.05$---at the expense of a large departure from the nominally energy-conserving line $\alpha = (3-k)/2$. We note that this departure is fairly general: adding $k$ to the fitted parameters for different fixed values of $\epsilon_e$ also leads to a model with strong deceleration into a stratified medium. More robust fitting is needed to determine whether this improved fit is an unphysical quirk of the model or a true effect. 

The green fixed $R$ effective LOS model makes a similar prediction to the first panel ($\pvnot = 0.63$, $n_0 = 10.97\hspace{2pt}\rm cm^{-3}$, $p=2.47$), but the change to $\epsilon_e$ greatly changes the accuracy of the fit. This discrepancy demonstrates the connection between $\epsilon_e$ and the shock dynamics -- keeping the other parameters fixed, the radius assumed in the fixed $R$ case is not necessarily the radius that fits the data best for small $\epsilon_e$. Choosing the radius to be the value of $R_0$ found from the full-volume calculation similarly leads to poor fits to the data. On the other hand, the fitted $R$ model fit closely resembles that of the full-volume model, but it makes a substantively different prediction ($\pvnot = 0.96$, $n_0 = 2.62\hspace{2pt}\rm cm^{-3}$, $p=2.51$, $R = 9.93\times10^{16}\hspace{2pt} \rm cm$) from both the full-volume model and the fitted $R$ model in the first panel. Note that in this case as well, the radius predicted by the fitted $R$ model is different from that in the full-volume value of $1.64\times10^{17}\hspace{2pt} \rm cm$ and the fixed $R$ value of $ 1.34\times10^{17}\hspace{2pt} \rm cm$). 

Figure~\ref{fig:CSS161010_comparison} demonstrates that at least two distinct physical models fit the data for CSS161010 rather well. This degeneracy is emblematic of the poorly constrained post-shock microphysics; in this case, the data cannot distinguish between different sets of microphysical assumptions.  Future work is needed to more systematically explore whether or not thermal electrons are important for modeling of FBOTs, and to check for consistency across temporal epochs.

If we treat the SED from the full-volume model as ``data" and fit the LOS approximation to it, leaving $R$ as a free variable, it is possible to recover the full-volume model for low $\pvnot$ as long as $\alpha=0$. In realistic events we expect some degree of deceleration, for which this simple extension fails. Moreover, the fit only works for low $\pvnot$ and, as noted above in the discussion of the fitted $R$ model, the physical meaning of the predicted radius is unclear. It is possible that the effective LOS approximation can be tuned in a more complicated way to match the full-volume model, but we leave further consideration of this issue as well as a more detailed, large-scale fitting of the full-volume model to future work. 

From the examples given in Figure~\ref{fig:CSS161010_comparison}, we conclude that including the radius in the effective LOS modeling is important, and that a simple choice of $R$ will fail in generic circumstances. On the other hand, even if we do fit for $R$, the parameters predicted by the effective LOS approximation do not match those of the full-volume calculation. Additionally, the physical meaning of the fitted radius is unclear for mildly relativistic shocks. Although the resulting curves may be similar with certain assumptions, the predicted model parameters vary widely due mainly to the presence of nonzero deceleration. If we constrain $\alpha$ to be 0 (not pictured here), the effective LOS approximations match the full-volume model more closely, but the model does not fit the data well. In general, a simple implementation of the effective LOS approximation cannot account for the significant effect of deceleration on the distribution of emitting radii.

\section{Summary and Discussion}\label{sec:discussion}

In this work, we have presented a numerical model which solves the full radiative transfer problem for synchrotron-emitting shocks,
accounting for relativistic effects without any approximations regarding the radiative transfer.
This `full-volume' model extends previous treatments and is specifically designed to be applicable to trans-relativistic explosions, where existing models are highly inaccurate (note that our full-volume model is accurate for any velocity, including non-relativistic and ultra-relativistic cases).
In particular, our model extends analytic approaches which typically employ an `effective LOS approximation' (e.g., \citealt{Chevalier98,Giannios&Metzger11,Margalit&Quataert21,MQ24}; see \S\ref{sec:approximations} for further discussion). It instead more closely follows the numerical formalism developed for GRB afterglows by \cite{Granot+99a,Granot+99b} while extending it in important ways (e.g., extending the treatment to be applicable also at low velocities; employing a flexible prescription for the hydrodynamics; including a treatment of thermal electrons).
We have also derived a semi-analytic `thin-shell' approximation which can be viewed as a ``half-way'' approach between simplified analytic models and the complete numerical treatment of the full-volume method.

The predictions of our full-volume calculations differ in important ways from previous approximate models. 
Although approximate analytic models are reasonably accurate when the shock velocity is non-relativistic, they noticeably diverge from the full calculation even at mildly relativistic velocities, $\pvnot \gtrsim 0.1$  (where $\pvnot$ is a characteristic value for the shock proper-velocity). For trans-relativistic shocks where $\pvnot \sim 1$, approximate methods can under-predict the true flux by over an order of magnitude,
particularly if the shock decelerates as a function of time or runs into a variable density CSM
(see Figure~\ref{fig:SED}). This finding is also true for ultra-relativistic shocks, for which $\pvnot \gg 1$.
These results imply that applying existing approximate models to trans-relativistic transients may produce wildly inaccurate inferences.
As a test case, we have briefly examined this issue by fitting different models to a single-epoch SED of the FBOT CSS161010 (\citealt{Coppejans20}; \S\ref{sec:curve_fit}). We find that the inferred shock velocities and ambient densities differ appreciably depending on whether the full-volume calculations or effective LOS models are employed (Figure~\ref{fig:CSS161010_comparison}). Implications will be explored more thoroughly in future work.

We investigate causes for the inaccuracy of the simple approximations in the optically thin part of the spectrum using  `emission region' contour plots (Figure~\ref{fig:contours}). These plots demonstrate that the emergent flux can be dominated by emission from regions far behind the shock front, especially in cases where the shock decelerates ($\alpha > 0$; Equation~\ref{prop_vel}) or the ambient CSM density drops as a function of distance from the origin of the explosion ($k>0$; Equation~\ref{n_ext}).
In certain trans-relativistic cases, emission from the the distal region (region pointing away from the observer, corresponding to $y<0$) can even contribute more to the total emission than points in the proximal region (nearer to the observer; Figure~\ref{fig:emission_ratio}).
These facts render the approximate effective LOS and thin shell models inaccurate as they cannot properly capture the three-dimensional spatial distribution of emitting matter.
Note, however, that the thin-shell approximation is always accurate for optically-thick frequencies, as in this case emergent emission does originate very close to the shock surface (e.g., Figure~\ref{fig:high_low_contour}).

More broadly, photon time-of-flight effects cannot be ignored for relativistic shocks. These effects cause photons that are detected contemporaneously at observed time $T$ to have been emitted at various different retarded times $t$. If the shock decelerates and/or runs into a stratified CSM, photons originating at different times $t$ were emitted when the shock conditions (e.g., the velocity or density) were different.
Consequently, there is no single shock radius or velocity that describe from emission a relativistic shocks at some epoch $T$. Instead, observed emission probes a range of shock radii, at which the shock velocity and external CSM density may have been different.
This is an important conceptual difference between the exact full-volume calculation and approximate analytic models, and highlights the shortcomings of the latter.
To explore this issue more quantitatively we have examined in Figure~\ref{fig:r_vals} the distribution of shock radii probed by contemporaneously observed emission as a function of shock velocity.
The plot shows that for relativistic explosions $\pvnot \gtrsim 1$, a wide range of shock radii are probed even in a single-epoch observation. Only exact calculations like our full-volume model can take this effect into account.

The formalism developed in this work is suited to the modeling of shocks in various settings. Of particular interest is the fitting of observed trans-relativistic sources such as FBOTs, jetted TDEs, NS mergers, and broad-line Type Ic supernovae. The full-volume model allows for a precise calculation of the physical parameters of these events and a detailed exploration of the possible effects of thermal electrons on their spectra and light curves. The full-volume model is also relevant to the physics of NS mergers, where it could be used to constrain the physics of GW170817 and to explore the observational signatures of magnetar-boosted synchrotron emission \citep[e.g.,][]{Metzger14,Schroeder20,Sarin22}. 
Our model is also valid for ultra-relativistic explosions, and can therefore be applied to GRB afterglows, 
although there may not be much added benefit in this particular context given that the GRB literature has evolved beyond this point \citep[e.g.,][]{Granot+99a,Granot+99b,GS02,vanEerten12,Ryan20,Wang24}. As a proof of concept and code validation test we have compared the results of our full-volume calculations to those of \cite{GS02} in the ultra-relativistic GRB setting, and find excellent agreement between the two models (Appendix~\ref{sec:Appendix_GRB}).
A separate setting where ultra-relativistic shocks could be involved and in which our new code may produce novel insights is fast radio bursts (FRBs). Specifically,  
the synchrotron maser model for FRBs invokes ultra-relativistic magnetized shocks, and predicts that FRBs should be accompanied by an incoherent synchrotron afterglow \citep[e.g.,][]{Lyubarsky14,Beloborodov17,Beloborodov,MargalitMetzger20,MargalitBeniamini20}. More accurate predictions for the signatures of such FRB counterparts 
can be calculated using our new code, and will be explored in future work.

The formalism described in this paper improves on many simpler approximations commonly used in the literature, but it does have limitations. 
Our current approach assumes spherical symmetry, which may be a reasonable first-order approximation for many non-relativistic (and perhaps some trans-relativistic) explosions, but ultra-relativistic explosions are instead often thought to be highly collimated and anisotropic. Even so, relativistic beaming ensures that the signatures of jetted ultra-relativistic outflows appear similar to emission from a spherical outflow so long as the outflow Lorentz factor is $\Gamma \gtrsim 1/\theta_{\rm j}$, where $\theta_{\rm j}$ is the jet opening angle (i.e. at times prior to the jet-break time; \citealt{Sari+99}).

In this present paper we have also adopted a simplistic  toy model for the shock dynamics and post-shock hydrodynamic profiles. In order to investigate the effects of deceleration, we have parameterized the shock dynamics using a power-law prescription (Equation~\ref{prop_vel}; see also Appendix~\ref{app:shock_radius}). This flexible prescription has enabled a comprehensive parameter survey using the deceleration parameter $\alpha$. It is also convenient because it reduces to a coasting solution when $\alpha=0$, and to the Sedov-Taylor (Blandford-McKee) self-similar energy-conserving solutions in the non-relativistic (ultra-relativistic) regimes when $\alpha = (3-k)/2$ (where $k$ is the power-law index of the CSM density profile; Equation~\ref{n_ext}).
Real explosions are expected to transition between these two extremes and may therefore not follow a single power-law prescription for the entirety of their evolution. Nevertheless, such a prescription may still be useful if one instead thinks of the deceleration parameter as an approximation of the local shock dynamics at some epoch, $\alpha = - d\ln \pv / d\ln R$.
For the purpose of simplicity, we have also assumed here that the post-shock hydrodynamic variables are spatially constant behind the shock so that the number density, energy density, and fluid velocity, are fully determined by the shock jump conditions (\S\ref{sec:downstream}).
This has the benefit of allowing us to compare the impact of relativistic effects between the full-volume model and effective LOS approximation. More realistic radial profiles for any of these quantities could easily be adopted instead, but any specific choice would likely depend on the relevant astrophysical context. This may be important when modeling observations in greater detail, but has limited relevance to the findings of our present work.
In any case, our formalism and code are not tied to these specific assumptions, and both can be used with arbitrary shock expansion histories and post-shock hydrodynamic profiles specified by the user.

In the present formulation, we have also neglected the effects of synchrotron cooling and inverse-Compton scattering on the post-shock electron distribution function. These effects can be very important in many cases, and have only been neglected here for the sake of simplicity. Although potentially important for accurate modeling of real-world data, these effects do not impact the overall conclusions of our present study.
Future work will incorporate these ingredients into our formalism, and further explore the implications of applying these accurate relativistic calculations to observed events.

Finally, although our canonical model includes a treatment of synchrotron emission and absorption from thermal electrons \citep[following][]{Margalit&Quataert21}, we note that the full-volume formalism presented here does not depend on this assumption. If desired, thermal electrons can be ``turned off'' within our existing code and framework. The underlying conclusions of our study, and the fundamental importance of treating relativistic explosions with full-volume methods rather than approximate one-zone models holds regardless of whether thermal electrons are important or not (an issue that is currently openly debated). 

The code described in this work is publicly available at \dataset[synchrotron\_shock\_model]{https://doi.org/10.5281/zenodo.17904998} and we encourage its use in future studies of relativistic synchrotron-emitting sources.

\begin{acknowledgments}
B.M. and R.F. 
are supported in part by the National Science Foundation under grant number AST-2508620. This work utilized of the NASA Astrophysics Data System. Software citation information was aggregated using \texttt{\href{https://www.tomwagg.com/software-citation-station/}{The Software Citation Station}} \citep{software-citation-station-paper, software-citation-station-zenodo}.
\end{acknowledgments}

\software{\texttt{matplotlib } \citep{Hunter:2007}, \texttt{numpy} \citep{numpy}, \texttt{python} \citep{python}, and \texttt{scipy} \citep{2020SciPy-NMeth, scipy_15716342}. Gemini (\citealt{Google_Gemini}; used exclusively for code-assist purposes related to aesthetic improvements to Figures~\ref{fig:high_low_contour}--\ref{fig:BM}).
}

\bibliography{refs}{}
\bibliographystyle{aasjournal}


\appendix

\section{Thin Shell Approximation}\label{appendix:thin_shell}

As described in \S\ref{sec:approximations}, the semi-analytic thin shell approximation assumes that all emission and absorption is sourced from an infinitely thin shell at the shock front. Taking into account photon time-of-flight effects that are relevant when the shock velocity approaches the speed of light (as described in \S\ref{model:kinematics}), the thin-shell approach implies that emission and absorption occur only along a surface defined by the EATS (Figure~\ref{fig:EATS}). The location of this surface is therefore an important ingredient for the thin-shell calculation. For a given coordinate $0 \leq x \leq 1$ in the direction perpendicular to the LOS, 
there are two y-coordinates that define the EATS,  $y_{{\rm back}}(x)$, and $y_{\rm front}(x)$, which can be found from the two solutions to Equation~(\ref{R_max1}) at any fixed x-coordinate. $y_{{\rm back}}(x)$ and $y_{\rm front}(x)$ represent the minimum and maximum of these two solutions, respectively (note that at $x=1$ the two solutions converge and there is only one solution).

The thin shell approximation implies that the emissivity and absorption coefficients can be written as

\begin{equation}
    j_\nu(r) = j_{\nu,0} R_\ell \delta \left( r-R \right) 
    ,~~~~~
    \alpha_\nu(r) = \alpha_{\nu,0} R_\ell \delta \left( r-R \right)
    .
\end{equation}
The radiative transfer equation along the LOS direction can be recast as an ODE as a function of the y-coordinate. At any given value of $x$ along which this equation will be solved, the radial coordinate can be thought of as a function of $y$, so that $r=r(y;x)$.
Using the delta-function identity $\delta[f(y)] = \delta(y-y_0)/|f'(y_0)|$ where $y_0$ is a zero of the function $f(y_0)=0$, and recognizing the relevant zero-points $y_0$ in our problem as $y_{\rm back}$ and $y_{\rm front}$, we have
 
$\delta[ r(y;x)-R ] = \mu(y_0) \delta[ y-y_0(x) ] / R_\ell$. Therefore
\begin{equation}\label{eq:thin_shell_coeffs}
    j_\nu(x,y) = j_{\nu,0} \mu\left[y_0(x)\right] \delta\left[ y-y_0(x) \right]
    ,~~~~~
    \alpha_\nu(x,y) = \alpha_{\nu,0} \mu\left[y_0(x)\right] \delta\left[ y-y_0(x) \right]
    .
\end{equation}

For the thin shell approximation to make sense, we must be careful in how the the number density in the thin shell is chosen. 
The density plays two different important roles. First, it controls the total number of emitting / absorbing electrons. This enters through the linear dependence of both emission and absorption coefficients on the density, $j_\nu, \alpha_\nu \propto n$. Second, the downstream number density and the corresponding energy-density govern the post-shock magnetic field $B$ and electron temperature $\Theta$ (see \S\ref{sec:downstream}).
In our current formulation of the thin-shell approximation we calculate the values of $B$ and $\Theta$ using the density determined by the shock-jump conditions, $n = 4\gamma_f n_{\rm ext}$.
On the other hand, we chose the density that goes directly into the emission and absorption coefficients in such a way that ensures that the total number of electrons within the shell equals the number of electrons swept-up by the shock. Defining the number density in the thin shell to be $n(r) \equiv n_0 R \delta(r-R) / 3$ and following a similar calculation as in Appendix~\ref{app:xi_shell} below we find that 
\begin{equation}
   n_0 = \frac{3}{3-k} n_{\rm ext}(R),
\end{equation}
where 
$n_{\rm ext}$ is the upstream number density, $R = R(x,y)$ is the shell radius, and we have assumed that $n_{\rm ext}\propto r^{-k}$ (Equation~\ref{n_ext}). 
The emission and absorption coefficients $j_0$ and $\alpha_0$ can then be calculated as in \S~\ref{subsec:rad_transfer}, but evaluated only at the EATS ($\xi=1$) and using the number density $3 \hspace{1pt} n_{\rm ext}/(3-k)$. Formally solving the radiative transfer equation $dI_\nu/ds = j_\nu -\alpha_\nu I_\nu$ in the observer frame, we find (integrating over the Dirac delta functions in Equation~\ref{eq:thin_shell_coeffs})

\begin{equation}
   I_\nu = \frac{j_0}{\alpha_0}\left(1-e^{-\alpha_0 \mu R_\ell }\right).
\end{equation}

We now apply this general solution to the thin shell geometry The intensity contributed by the back side of the shock (at coordinate $y_0 = y_{\rm back}$) is 
\begin{equation}\label{thin_shell_back}
I_{\nu,{\rm back}}(x) = D^3\frac{j'_\nu}{\alpha'_\nu}\left(1-e^{-\alpha'_\nu\mu R_\ell D^{-1}}\right)\bigg|_{x,y_{\rm_{\rm back}}(x)}
.
\end{equation}
The final emergent intensity $I^{\rm thin}_{\nu}$ at the front end of the shock consists of a similar solution at $y_0 = y_{\rm front}$, plus a term containing an absorbed $I_{\nu,{\rm back}}$ from the back part of the EATS. Overall, the total emergent intensity is therefore
\begin{flalign}
I_{\nu}(x) & = D^3\frac{j'_\nu}{\alpha'_\nu}\left(1-e^{-\alpha'_\nu\mu R_\ell D^{-1}}\right)\bigg|_{x,y_{\rm front}(x)} + I_{\nu,{\rm back}}(x) e^{-\alpha'_\nu\mu R_\ell D^{-1}}\bigg|_{x,y_{\rm front}(x)}
.
\end{flalign}
With $I_\nu(x)$ in hand, we can numerically integrate Equation~(\ref{F_nu}) to calculate the flux in the thin shell approximation.

\section{Model of Shock Dynamics with Power-law Deceleration}\label{app:shock_radius}

To derive the expansion history of the shock radius $R(t)$, it is useful to first consider the simple case when the shock front moves at a constant velocity $R = \beta_{\rm sh}ct$. For an expansion of this form
we can use the relation $T= t-r\mu/c$ at the shock front ($r=R$) to obtain an expression for the retarded time $t = T / (1-\beta_{\rm sh} \mu)$.

The equal-arrival-time surface (EATS) for such a constant-velocity shock (corresponding to $\alpha=0$; Equation~\ref{prop_vel}) is therefore
\begin{equation}\label{const_vel}
    R^{(\alpha=0)}_{\rm EATS}(\mu, T) 
    = \frac{\beta_{\rm sh}c T}{1-\beta_{\rm sh}\mu} .
\end{equation}
This describes an ellipse with one of the foci centered on the origin. The corresponding eccentricity and semi-major axis of this ellipse are $e = \beta_{\rm sh}$ and $a = \beta_{\rm sh} c T / (1-\beta_{\rm sh}^2) = \Gamma_{\rm sh}^2 \beta_{\rm sh} cT$, respectively.

Relaxing the assumption of constant velocity, we consider the shock radius measured at the LOS $R_\ell$ ($\mu=1$) at an observed time $T$. During an infinitesimal time interval $dT$ along the LOS, any shock can be considered to be moving at a constant velocity. Thus we may write
\begin{equation} \label{eq:Appendix_dTdR}
c \,dT = \left. \frac{1-\beta_{\rm sh}}{\beta_{\rm sh} c} \right\vert_{R=R_\ell} dR_\ell 
= \left( \sqrt{1+\pvell^{-2}} - 1 \right) dR_\ell
,
\end{equation}
where in the second equality we have rewritten $\beta_{\rm sh}$ in terms of the shock proper-velocity $\pvell = \left. \beta_{\rm sh}/\sqrt{1-\beta_{\rm sh}^2} \right\vert_{R=R_\ell}$
at the LOS radius $R=R_\ell$.

For the power-law evolution assumed in this paper $\pv \propto R^{-\alpha}$ (Equation~\ref{prop_vel}). Using this choice and changing variables to $z = R_\ell g^{-1/\alpha}$, with the constant $g = \pvnot R_0^{\alpha}$, we can integrate both sides of Equation~(\ref{eq:Appendix_dTdR}) and obtain
\begin{equation}
c T 
= g^{1/\alpha}\int \left(\sqrt{1+z^{2\alpha}}-1\right)dz\\
= g^{1/\alpha} z\left\{_2F_1\left(-\frac{1}{2},\frac{1}{2\alpha},\frac{2\alpha+1}{2\alpha},-z^{2\alpha}\right)-1\right\}
,
\end{equation}
where $_2 F_1$ is the hypergeometric function.
Since $z^{2\alpha} = \pvell^{-2}$
and $\pvell = \pvnot \left(R_\ell/R_0\right)^{-\alpha}$,
\begin{equation}\label{RLOS}
cT = R_\ell \left\{_2F_1\left(-\frac{1}{2},\frac{1}{2\alpha},\frac{2\alpha+1}{2\alpha},-\pvnot^{-2} \left(\frac{R_\ell}{R_0}\right)^{2\alpha}\right)-1\right\} .
\end{equation}
Rewriting this in terms of the retarded time $t_\ell = T + R_\ell/c$, we have 
\begin{equation}\label{RLOS2}
ct_\ell = R(t_\ell) _2F_1\left(-\frac{1}{2},\frac{1}{2\alpha},\frac{2\alpha+1}{2\alpha},-\pvnot^{-2} \left(\frac{R(t_\ell)}{R_0}\right)^{2\alpha}\right) .
\end{equation}
Equation~(\ref{RLOS2}) implicitly defines the function $R(t)$ in the shock frame for all retarded times $t$. In other words, Equation~(\ref{RLOS2}) is an implicit solution to the shock radius as a function of time, $R(t)$, assuming the power-law deceleration model defined via Equation~(\ref{prop_vel}). Thus, we can now use this equation to define the EATS radius $R_{\rm EATS}(\mu,T)$ seen by an observer by calling $R(t)$ given above at the retarded time $t = T + \mu R_{\rm EATS}/c$,
\begin{equation}\label{R}
cT + \mu R_{\rm EATS}(T,\mu) = R_{\rm EATS}(T,\mu) \left\{_2F_1\left(-\frac{1}{2},\frac{1}{2\alpha},\frac{2\alpha+1}{2\alpha},-\pvnot^{-2} \left(\frac{R_{\rm EATS}(T,\mu)}{R_0}\right)^{2\alpha}\right)\right\} .
\end{equation}
This implicit equation provides a complete solution for $R_{\rm EATS}(T,\mu)$ as a function of the characteristic proper-velocity $\pvnot$ and the shock deceleration parameter $\alpha$. This equation must generally be solved numerically; however, analytic expressions can be obtained for certain cases of interest. Specifically, using Equation~(\ref{R}), we can obtain more explicit values for $R_\ell$, $R_0$, and $X_{\perp}$. In the limit of large $\pvell$, $R_\ell \equiv R_{\rm EATS}(T,\mu=1)$ satisfies
\begin{equation}\label{RL1}
cT = R_\ell \left\{_2F_1\left(-\frac{1}{2},\frac{1}{2\alpha},\frac{2\alpha+1}{2\alpha},-\pvell^{-2}\right) -1\right\} 
\overset{\pvell \gg 1}{\approx} 
R_\ell \left\{1 + \frac{1}{4\alpha+2}\pvell^{-2} +\ldots-1\right\}.
\end{equation}
In the non-relativistic limit $\pvell \ll 1$, we have instead
\begin{equation}\label{RL2}
cT 
\overset{\pvell \ll 1}{\approx} 
R_\ell \left\{\frac{2\alpha \Gamma( \frac{2\alpha+1}{2\alpha})}{(\alpha+1)\Gamma(1/2\alpha)} \pvell^{-1} + \frac{ \Gamma( \frac{-\alpha+1}{2\alpha})\Gamma( \frac{2\alpha+1}{2\alpha})}{2\sqrt{\pi}}\pvell^{1/\alpha} + O(\pvell)\right\},
\end{equation}
where $\Gamma(x)$ is the Gamma function. For $\alpha>0$, the first term on the right dominates. Putting Equations~(\ref{RL1},\ref{RL2}) together, we have 
\begin{equation}\label{RL}
    R_\ell
    \simeq
    \begin{cases}
        \frac{(\alpha+1)\Gamma(1/2\alpha)} {2\alpha \Gamma( \frac{2\alpha+1}{2\alpha})} \pvell cT&,~~ \pvell \ll 1
        \\
        (4\alpha+2) \pvell^2 c T &,~~ \pvell \gg 1
    \end{cases}
\end{equation}
Note that $\pvell  = \pvnot \left( R_\ell/R_0 \right)^{-\alpha}$, so that Equation~(\ref{RL}) defines $R_\ell$ implicitly in each limit.

The maximum distance perpendicular to the LOS is reached at $X_\perp = \max_{\mathbf{\mu}} \hspace{2pt} \left[ \sqrt{1-\mu^2} R_{\rm EATS}(\mu, T) \right]$, which is defined to occur at angle $\mu = \mu_\perp$ and physical radius $R_0 = R_{\rm EATS}(\mu=\mu_\perp, T)$. 
The angle $\mu_\perp$ reached at $R = R_0$ satisfies 
\begin{equation}\label{deriv}
\frac{d}{d\mu}\left[ \sqrt{1-\mu^2} R_{\rm EATS} \right]_{\mu=\mu_\perp} = 0 .
\end{equation}

From Equation~(\ref{R}), we have 
\begin{equation}\label{mu_2F1}
\mu(R_{\rm EATS},T) = \hspace{0.00001pt}_2F_1\left(-\frac{1}{2},\frac{1}{2\alpha},\frac{2\alpha+1}{2\alpha},-\pv^{-2} \right) - \frac{cT}{R_{\rm EATS}}.
\end{equation}
Taking the derivative of Equation~(\ref{mu_2F1}) and using hypergeometric function identities, we obtain 
\begin{equation}
\frac{d\mu}{dR_{\rm EATS}} = \frac{1}{R_{\rm EATS}}\left\{\sqrt{1+\pv^{-2}} - \hspace{0.01pt}_2F_1\left(-\frac{1}{2},\frac{1}{2\alpha},\frac{2\alpha+1}{2\alpha},-\pv^{-2} \right)  + \frac{cT}{R_{\rm EATS}}\right\} = \frac{\sqrt{1+\pv^{-2}} - \mu}{R_{\rm EATS}}.
\end{equation}

Combining the equations above and noting that $\left. \left(\Gamma\beta\right)_{\rm sh} \right\vert_{R=R_0} = \pvnot$ yields

\begin{equation}\label{eq:mu_perp}
    \mu_\perp = \frac{1}{\sqrt{1+\pvnot^{-2}}} = \beta_{\rm sh,0}.
\end{equation}
Using Equation~(\ref{R}) and the definition $R_0 = R_{\rm EATS}(\mu_\perp)$, we arrive at a closed-form expression for $R_0$,
\begin{equation}\label{R0}
R_0 = \frac{cT}{ _2F_1\left(-\frac{1}{2},\frac{1}{2\alpha},\frac{2\alpha+1}{2\alpha},-\pvnot^{-2} \right)- \left[ 1+\pvnot^{-2} \right]^{-1/2}}  .
\end{equation}
Using the same asymptotic expansions for $_2F_1$ as above, this reduces to
\begin{equation}\label{R0_lims}
    R_0
    \simeq
    \begin{cases}
        \frac{(\alpha+1)\Gamma(1/2\alpha)} {2\alpha \Gamma( \frac{2\alpha+1}{2\alpha})} \pvnot cT&,~~ \pvnot \ll 1
        \\
        \frac{2\alpha+1}{\alpha+1} \pvnot^2 c T &,~~ \pvnot \gg 1
    \end{cases}
\end{equation}
in the non-relativistic and ultra-relativistic regimes.
For $\alpha = 3/2$ these limits correspond to the Sedov-Taylor and Blandford-McKee solutions, respectively (see also discussion in \S\ref{sec:downstream}).
With, the expressions above, we show that the ratio of $R_\ell$ to $R_0$ in simply an order-unity factor,
\begin{equation}\label{RlR0}
    \frac{R_\ell}{R_0}
    \simeq
  \begin{cases}
        1 &,~~ \pvnot \ll 1
        \\
        \left[ 2 (\alpha+1) \right]^{1/(2\alpha+1)} &,~~ \pvnot \gg 1
    \end{cases}
\end{equation}

Finally, we obtain estimates for the minimum EATS radius $|y_{\rm min}| R_\ell$ using the same procedure as above. 
The smallest EATS radius always occurs at the very back of the emitting regions, when $\mu = -1$ (see Figure~\ref{fig:EATS}).
The non-relativistic limit is trivial, $|y_{\rm min}| \simeq 1$ when $\pvnot \ll 1$.
In the ultra-relativistic limit we find 
to leading order that $|y_{\rm min}| R_\ell \simeq cT/2$. Combining these results yields
\begin{equation}\label{y_min_scaling}
   |y_{\rm min}|R_\ell
    \simeq
    \begin{cases}
        \frac{(\alpha+1)\Gamma(1/2\alpha)} {2\alpha \Gamma( \frac{2\alpha+1}{2\alpha})} \pv cT&,~~ \pvnot \ll 1
        \\
        \frac{cT}{2} &,~~ \pvnot \gg 1
    \end{cases}
\end{equation}
Note that the ultra-relativistic limits in Equations~(\ref{R0}, \ref{RlR0}, \ref{y_min_scaling}) imply that $|y_{\rm min}| \propto \pvnot^{-2}$ when $\pvnot \gg 1$. Therefore the minimum EATS radius $|y_{\rm min}|R_\ell$ is significantly smaller than $R_\ell$ or $R_0$ when the shock is ultra-relativistic.

\section{Calculation of Emitting Region}\label{app:xi_shell}
In this appendix, we calculate the size of the emitting region. We work here in the rest frame of the un-shocked CSM, and assume a self-similar post-shock density profile. Imposing particle number conservation, the downstream density $n$ and the upstream density $n_{\rm ext}$ must satisfy
\begin{equation}\label{}
\int_0^{R_{\rm sh}(t)} n(r,t) 4\pi r^2 dr = \int_0^{R_{\rm sh}(t)} n_{\rm ext}(r) 4\pi r^2 dr ,
\end{equation}
subject to the boundary condition $n(r=R_{\rm sh},t) = 4\gamma_f^2(r=R_{\rm sh},t)
\hspace{1pt} n_{\rm ext}(R_{\rm sh})$, as implied by the shock-jump conditions \citep[e.g.,][]{BlandfordMcKee76}. Due to the assumption of self-similarity, we can re-write the density as 
\begin{equation}
    n(r,t) = \tilde{n}(t) h(\xi) ,
\end{equation}
with $\xi = r/R_{\rm sh}$ the self-similar coordinate. Without loss of generality we choose $h(1) = 1$. The boundary condition then implies that $\tilde{n}(t) =4 \gamma_f(R_{\rm sh}[t],t)^2 n_{\rm ext}(R_{\rm sh}[t])$. Making this substitution in the integrals, the radial dependence of the post-shock number density $h(\xi)$ satisfies the condition
\begin{equation}\label{eq:h_xi_integral}
\int_0^1 h(\xi) \xi^2 d\xi = \frac{1}{4 \gamma_f^2}\int_0^1 \frac{n_{\rm ext}(\xi R_{\rm sh})}{n_{\rm ext}(R_{\rm sh})} \xi^2 d\xi .
\end{equation}

For a power-law external CSM density $n_{\rm ext}(r) = n_0(r/R_0)^{-k}$, the right-hand side can be integrated to give
$1 / [ 4 \gamma_f^2(3-k) ]$.

In  the simple model described in Section 3, we assume that
\begin{equation}
h(\xi) = \bigg\{
\begin{array}{ll}
      1 & \xi_{\rm shell}\leq \xi \leq 1 \\
      0 & \xi<\xi_{\rm shell}, \xi>1 .\\
\end{array} 
\end{equation}

Enforcing particle conservation (Equation~\ref{eq:h_xi_integral}) then implies that
\begin{equation}\label{xi_shell}
\xi_{\rm shell} = \bigg(1- \frac{3}{3-k}\frac{1}{4\gamma_f^2(t)}\bigg)^{1/3}.
\end{equation}

Putting this all together, the density in the local comoving fluid rest frame $n' = n/\gamma_f$ can be written as
\begin{equation}\label{n_prime1}
n'(r,t) = 4n_0 \frac{\gamma_f(R_{\rm sh},t)^2}{\gamma_f(r,t)} \bigg(\frac{R}{R_0}\bigg)^{-k}
,
\end{equation}
which, in the case where $\gamma_f$ is radially constant (as assumed in the simple model described above; see also \S\ref{sec:downstream}), 
reduces to 
\begin{equation}\label{n_prime2}
n'(r,t) = 4n_0 \gamma_f(t) \bigg(\frac{R}{R_0}\bigg)^{-k}.
\end{equation}

To compare the model described in this paper to the effective LOS approximation, we define a volume-filling factor $f$ following \cite{MQ24} (see also \citealt{Chevalier98}) so that $4 \pi f R^3 / 3$ is the total emitting volume
$4\pi ( R^3 - \xi_{\rm shell}^3 R^3) / 3$ .
This leads to the correspondence
\begin{equation}\label{eq:f}
    f = 1-\xi_{\rm shell}^3 
    = \frac{3}{3-k}\frac{1}{4\gamma_f^2}
    .
\end{equation}

\section{Comparison to GRB Afterglow Models}
\label{sec:Appendix_GRB}

In the ultra-relativistic limit, we compare the model developed in this paper to GRB afterglow models, which make use of the Blandford-McKee solution \citep{Sari98,Granot+99a,Granot+99b,GS02}. In this section, we ignore the effect of thermal electrons,  assume that the mean molecular weights are $\mu_u = \mu_e = 1$, relax the $\xi_{shell}$ cutoff of the emitting region, and change the minimum power-law electron Lorentz factor  $\gamma_{\rm min}$ to the standard value used in GRB modeling (\citealt{Granot+99a}; their Equation~5). We examine two models with different external density profiles, $k=0$ and $k=2$. Fixing $\alpha = (3-k)/2$ for an energy-conserving shock, we use the formulas for the energy density $u'$, post-shock electron number density $n_e'$, and post-shock electron Lorentz factor $\gamma$ given by the Blandford-McKee solution (e.g., Equation~2 in \citealt{Granot+99a}). In the case common in ultra-relativistic shocks where the synchrotron frequency $\nu_{\rm m}$ of an electron with Lorentz factor $\gamma_{\rm min}$ is greater than the self-absorption frequency $\nu_{\rm sa}$, the absorption coefficient used above must be divided by $3(p+2)/4$ due to 
a discontinuity in the power-law distribution at $\gamma = \gamma_{\rm min}$, a discrepancy noted by \cite{GS02} (see Footnote~6 in that work). \citealt{GS02} include two types of cooling that affect the electron distribution function: synchrotron cooling (due to radiative losses) and adiabatic cooling. We neglect synchrotron cooling in this appendix, and artificially remove the associated cooling-break from the \cite{GS02} expressions for the purpose of this comparison.  Although our model does not explicitly incorporate adiabatic cooling of injected electrons, our approach calculates the emissivity and absorption coefficients at any given point based on the local fluid properties (in particular the energy density and number density). For a Blandford-McKee solution and purely power-law electron distribution (as is the case in this comparison), this treatment leads to the same result as if electrons were instead injected with some initial distribution and allowed to undergo adiabatic cooling, following the formalism of \cite{GS02}.

\begin{figure*}
 \centering
  \includegraphics[width=0.6\textwidth]{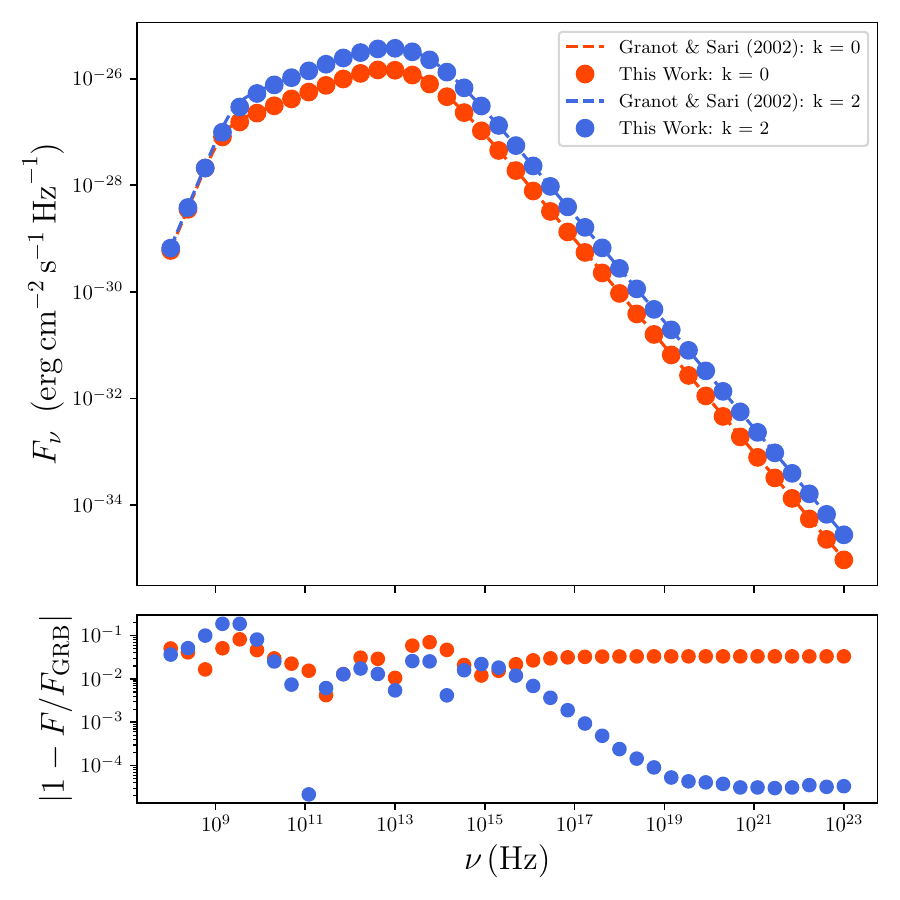}
  \caption{A comparison of the specific flux $F_\nu$ computed using Table 2 of \cite{GS02} and using the model developed in this paper and adapted for an ultra-relativistic Blandford-McKee solution. The agreement is typically at the several percent level, indicating that the code and formalism developed in our present work are able to replicate the relativistic radiative transfer calculations of \cite{Granot+99b}. This can be viewed as a test problem for validating our code. 
  }
  \label{fig:BM}
  \end{figure*}

For concreteness, we perform a comparison using the following parameters: $\pvnot = 100$, $n_0 = 10^{-3}\hspace{2pt}\rm cm^{-3}$, $p=3$, $\epsilon_B=0.1$, $\epsilon_e=0.01$, and $T=1\hspace{2pt}\rm hr$. 
These choices correspond to isotropic-equivalent explosion energies of $E_{\rm iso} = 1.14\times 10^{51} \,\rm erg$ and $E_{\rm iso} = 1.25\times 10^{51} \,\rm erg$ for the $k=0$ and $k=2$ cases (respectively).
In Figure~\ref{fig:BM}, we compare this model with the GRB fitting formula prescribed by \cite{GS02} (cases $b=1$ and $b=2$ in Table~2 of that work). \cite{GS02} note that these fitting formulas are accurate to within $7\%$ of numerical models calculating the full radiative transfer \citep{Granot+99b,GS02}. 
As seen in Figure~\ref{fig:BM}, the calculations resulting from our `full-volume' model are in very good agreement with the \cite{GS02} results, successfully replicating the GRB model. This agreement is typically within several percent level or less, with slightly higher disagreement (up to $19\%$) for the $k=2$ model near the self-absorption frequency $\nu_{\rm sa} \sim 10^{9}\,{\rm Hz}$. 

\end{document}